\documentclass[twocolumn,amsmath,amssymb,aps,prb,superscriptaddress]{revtex4-2}
\usepackage{epsf}
\usepackage{graphicx}
\usepackage{sidecap}
\usepackage{soul}
\usepackage{array}
\usepackage{amsmath}
\usepackage{amssymb}
\usepackage{dcolumn}
\usepackage{epstopdf}
\usepackage{bm}
\usepackage{amsmath}
\usepackage{physics}

\usepackage{gensymb}
\usepackage{color}
\usepackage{hyperref}
\usepackage{soul}
\sethlcolor{green}
\usepackage{bbding}
\usepackage{setspace}
\hypersetup{
    colorlinks=true,
    citecolor=red,
    linkcolor=red,
    filecolor=blue,   
    urlcolor=blue,
}

\usepackage[normalem]{ulem}

\usepackage{graphicx,color}
\usepackage{amsfonts}
\usepackage[figuresright]{rotating}  
\usepackage{amssymb}
\usepackage{amsmath}

\usepackage{bbold}
\usepackage{wrapfig,lipsum,booktabs}
\usepackage{mathtools}
\usepackage{psfrag}
\usepackage{subfigure}
\usepackage{multirow}
\usepackage{tabularx}
\usepackage{textcomp}
\usepackage{bm}
\usepackage{hyperref}
\usepackage{relsize}
\hypersetup{
 pdfnewwindow=true, colorlinks=true,
 linkcolor=blue, anchorcolor=blue,
 citecolor=blue, filecolor=blue,
 menucolor=blue, urlcolor=blue}

\def\beq{\begin{eqnarray}}
\def\eeq{\end{eqnarray}}

\usepackage[dvipsnames]{xcolor}

\def \beq {\begin{equation}}
\def \eeq {\end{equation}}
\def\bibsection{\refname}
\renewcommand{\refname}{\noindent\textbf{References}}

\begin{document}

\title{\Large Electronic Structure of a Nodal Line Semimetal Candidate TbSbTe}

\author{Iftakhar~Bin~Elius}\affiliation {Department of Physics, University of Central Florida, Orlando, Florida 32816, USA}
\author{Jacob~F~Casey}\affiliation{Department of Physics, SUNY Buffalo State, Buffalo, New York 14222, USA}
\author{Sabin~Regmi}\affiliation {Department of Physics, University of Central Florida, Orlando, Florida 32816, USA} \affiliation {Center for Quantum Actinide Science and Technology, Idaho National Laboratory, Idaho Falls, Idaho 83415, USA}
\author{Volodymyr~Buturlim}\affiliation {Glenn T. Seaborg Institute, Idaho National Laboratory, Idaho Falls, Idaho 83415, USA}
\author{Anup~Pradhan~Sakhya}\affiliation {Department of Physics, University of Central Florida, Orlando, Florida 32816, USA}
\author{Milo~Sprague}\affiliation {Department of Physics, University of Central Florida, Orlando, Florida 32816, USA}
\author{Mazharul~Islam~Mondal}\affiliation {Department of Physics, University of Central Florida, Orlando, Florida 32816, USA}
\author{Nathan~Valadez}\affiliation {Department of Physics, University of Central Florida, Orlando, Florida 32816, USA}
\author{Arun~K~Kumay}\affiliation {Department of Physics, University of Central Florida, Orlando, Florida 32816, USA}
\author{Justin~Scrivens}\affiliation {Department of Physics, University of Central Florida, Orlando, Florida 32816, USA}
\author{Yenugonda~Venkateswara}\affiliation {Department of Physics, SUNY Buffalo State, Buffalo, New York 14222, USA}
\author{Shovan~Dan}\affiliation {Institute of Low Temperature and Structure Research, Polish Academy of Sciences, ul. Ok\'olna 2, 50-422 Wroc{\l}aw, Poland}
\author{Tetiana~Romanova}\affiliation {Institute of Low Temperature and Structure Research, Polish Academy of Sciences, ul. Ok\'olna 2, 50-422 Wroc{\l}aw, Poland}
\author{Arjun~K~Pathak}\affiliation {Department of Physics, SUNY Buffalo State, Buffalo, New York 14222, USA}
\author{Krzysztof~Gofryk}\affiliation {Center for Quantum Actinide Science and Technology, Idaho National Laboratory, Idaho Falls, Idaho 83415, USA}
\author{Andrzej~Ptok}\affiliation {Institute of Nuclear Physics, Polish Academy of Sciences, W. E. Radzikowskiego 152, PL-31342 Krak\'ow, Poland}
\author{Dariusz~Kaczorowski}\affiliation {Institute of Low Temperature and Structure Research, Polish Academy of Sciences, ul. Ok\'olna 2, 50-422 Wroc{\l}aw, Poland}
\author{Madhab~Neupane}\thanks{Corresponding author:\href{mailto :madhab.neupane@ucf.edu}{~madhab.neupane@ucf.edu}}\affiliation {Department of Physics, University of Central Florida, Orlando, Florida 32816, USA}

\begin{abstract}
The $Ln$SbTe ($Ln$ = Lanthanides) family, like isostructural ZrSiS-type compounds, has emerged as a fertile playground for exploring the interaction of electronic correlations and magnetic ordering with the nodal line band topology. Here, we report a detailed electronic band structure investigation of TbSbTe, corroborated by electrical transport, thermodynamic, and magnetic studies. Temperature-dependent magnetic susceptibility and thermodynamic transport studies indicate the onset of antiferromagnetic ordering below $T_N \sim 5.8$~K. The electronic band structure study, carried out with high-resolution angle-resolved photoemission spectroscopy (ARPES) measurements aided with density functional theory based first-principles calculations reveal presence of a nonsymmorphic symmetry protected Dirac crossing in the $\Gamma$-X high symmetry direction, which is part of a nodal line along the X-R high symmetry direction. Another Dirac crossing occurs along the $\Gamma$-X direction at a relatively higher binding energy, which occurs from $\Tilde{\mathcal{C}}_{2\mathcal{\nu}}\mathcal{P}$ symmetry which is gapped in the theoretical calculations with the effect of spin orbit coupling considered. Parallel to this direction, our theoretical calculations and experimental results exhibit strongly momentum dependent surface bands. This study opens up an avenue to further uncover the intricate interplay among symmetry-protected topological band structure, spin-orbit coupling, and magnetism in this material and $Ln$SbTe family, in general.

\end{abstract}

\maketitle

\section{Introduction}

Since the discovery of topological insulators (TIs), the fascinating field of topological quantum materials (TQMs) has emerged as one of the newer frontiers of physics~\cite{Hasan_1, Qi_TI, MNreview, Sato_2017}. Serving as the arena for numerous foundational physical phenomena, these topological materials are continually evolving through robust theoretical and experimental investigations. This evolution has fostered the recognition of a variety of topological semimetals, including but not limited to Dirac~\cite{wang2, wang, Neupane2014, ZK_liu, Yang2014} and Weyl semimetals~\cite{Xu2015, Weylding, Soluyanov2015, Yan_felser_2017, Sakhya_2023_weyl}, nodal line semimetals (NLSMs)~\cite{node_surface_liang, Fang_NLSM_2015, MNzrsis, schoop2016dirac, bian2016topological, hosen_ZrSiX, nodal_loop, Armitagereview, weyl_nodal_loop} and Dirac nodal arcs~\cite{Wu2016_nodal_arc, gyanendra_ti2te2p}.
In a system with both time-reversal symmetry and inversion symmetry, band crossings can lead to the formation of Dirac nodes.
At a Dirac node, two doubly degenerate bands touch each other, resulting in a four-fold degenerate crossing point.
The linear dispersion relation around the Dirac node leads to massless Dirac fermions as the low-energy excitations in the system. We can classify or differentiate these TQMs based on the dimensionality of their band interactions or crossings in the momentum space~\cite{wang, ZK_liu}.
The idea of zero dimensional band contact points in Dirac or Weyl semimetals is further extended to higher dimensional nodal lines or surfaces~\cite{Chiu_2014, Fang_NLSM_2015, MNzrsis}. In NLSMs, the band interactions extend in forms of lines or closed loops, which are protected by extra symmetries like mirror reflection, inversion, time-reversal, spin-rotation or nonsymmorphic symmetries~\cite{Fang_NLSM_2015, MNzrsis, nonsymmomorphic_Yang, schoop2016dirac}. 
Nonsymmorphic symmetry incorporates a point group operation like fractional lattice translation with a nonprimitive lattice translation like screw axis or glide operation~\cite{nonsymmomorphic_zhao}. The pursuit of nonsymmorphic topological materials has been largely navigated by the seminal research throughout last few years, yet the number of material families featuring non-accidental nodal lines bereft of spin orbit coupling (SOC) is significantly limited~\cite{kane_theory_2016, schoop2016dirac, Schoop, nodaltheory, Toppfloating}. Some of the recent studies elucidated the possibility of continuous Dirac nodal points, under the condition that 2D square motifs twisted into hosting unit cell leading to a glide symmetry~\cite{kane_theory_2016, Schoop}. 
These theoretical speculations came into existence since such nodal line topological phases were observed in ZrSiS followed by other $MZX$ ($M$= Transition elements, $Z$= Si, Ge, Sb, Sn and $X$= S, Se, Te) materials with PbFCl type crystal structure~\cite{MNzrsis, schoop2016dirac}. When the transition metals are replaced by rare earth elements, different topological properties can be expected due to the correlations between 4$f$ and conduction band electrons and magnetism inherent to 4$f$ states within the $Ln$SbTe family. GdSbTe, one of the first materials reported from this family, has ZrSiS like nodal line characteristics, translated in energy positions~\cite{GdSbTe_hosen}. In HoSbTe, SOC gap opening was experimentally observed at X point~\cite{HoSbTe_transport, HoSbTe_shumiya}, turning it into a promising material for weak topological insulator. Also its band structure has been reported to show magnetism driven changes upon antiferromagnetic (AFM) to ferromagnetic (FM) transition near 4~K~\cite{HoSbTe_shumiya}. Similar metamagnetic transitions~\cite{CeSbTe_Lv}, charge density wave (CDW) state and weak Kondo effect, were observed in CeSbTe~\cite{CeSbTe_peng,CeSbTe_Lv, CeSbTe_cao}. In LaSbTe, at least two nodal lines along X-R and A-M directions were found to be robust against SOC protected by the nonsymmorphic symmetry~\cite{La, valadez2025low, binelius.ersbte.25}. Angle resolved photoemission spectroscopy (ARPES) studies on NdSbTe revealed two nodal lines along R-X and one diamond shaped nodal line structure around the $\Gamma$ point, as well as bands forming nodal lines along $\Gamma$-M direction~\cite{Nd111}. Magnetic and electronic study on NdSbTe revealed characteristic metamagnetic transitions in AFM region (below $T_N\sim$ 2.7~K), Kondo localization and enhanced electronic correlation~\cite{NdSBTe_pandey}. Experimental studies on SmSbTe showed multiple Dirac nodes which are part of nodal lines along $\Gamma$-X and X-R high symmetry (HS) directions~\cite{SmSbTe_sabin, SmSbTe_krishna_pandey}. Very recently, ARPES based studies on PrSbTe revealed similar electronic structure with gapless nodal line features~\cite{PrSbTe}. The Terbium variant of this series, TbSbTe remains underexplored with respect to the employment of experimental methodologies for the simultaneous investigation of its electronic structure and its electrical, magnetic, and thermodynamic properties. A preceding analysis utilizing neutron diffraction methods on TbSbTe brought to light the existence of competing magnetic phases within the AFM state~\cite{Tb111_neutron}. Complementary to this, another study on TbSbTe presented an amalgamation of thermodynamic, magnetic, and electronic transport measurements, supplemented by theoretical computations of the electronic band structure~\cite{Tb111}. Notwithstanding these contributions, an experimental exploration of TbSbTe's electronic band structure via ARPES with comprehensive integration with thermodynamic, electronic and magnetic measurements, to scrutinize the nodal line attributes of this variant remains unreported. This advancement is poised to significantly enrich our understanding of the unique quantum properties of this lesser-studied compound.\\
\indent In this paper, we present thermodynamic, magnetic, electrical transport and magnetoresistive studies of the TbSbTe single crystals. To investigate the electronic band structure of the material, we performed 
our electronic band structure studies on TbSbTe via ARPES, corroborated with the first principles calculations. The ARPES measurements were carried out at paramagnetic (PM) phase (at 18~K), that harbor persistent nodal line like features along several HS directions. Within the resolution of the ARPES spectra, we observe gapless band crossings along the $\overline{\Gamma}$-$\overline{\text{X}}$ and $\overline{\Gamma}$-$\overline{\text{M}}$ directions that form a nodal line along bulk X-R line and a diamond shape centered at the $\overline{\Gamma}$ point in $k_{x}-k_{y}$ plane, respectively. This research provides key insights into the electronic structure and topological states of TbSbTe, introducing a valuable foundation for understanding the topological properties of $Ln$SbTe family.
\begin{figure*}
	\centering
	\includegraphics[width= 18 cm]{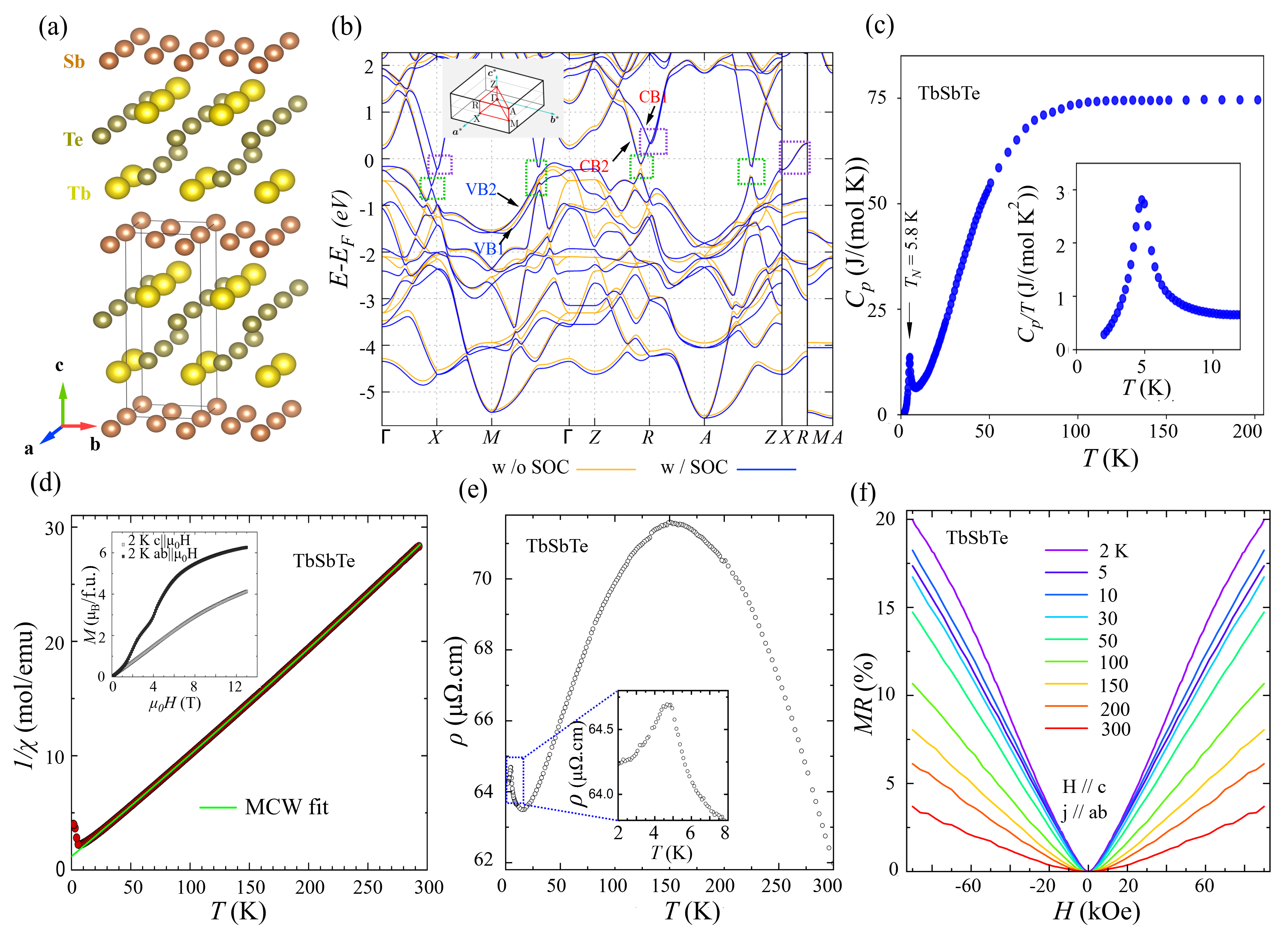}
	\caption{Crystal structure, electronic band structure, thermodynamic, magnetic and transport properties of TbSbTe. (a) Crystal structure of TbSbTe, two zig-zag chains of Tb-Te are sandwitched by square planar 2D Sb- nets. (b) Calculated band structure in paramagnetic phase, without considering SOC (yellow) and with consideration of SOC (blue). The positions of the nodal lines (which corresponds to the experimental observations) are marked with green boxes. The result corresponds to calculations where the $f$-states were treated as core states. (c) Temperature dependence of specific heat of TbSbTe, the black arrow indicates the AFM to PM phase transition, $C_p/T$ vs $T$ curve zoomed in at the low temperature range is presented in the inset. (d) Temperature dependence of the inverse dc susceptibility $\chi^{-1}$, measured in a magnetic field of, $\mu_{0}H= 0.5$~T. The solid green line represents the modified Curie--Weiss fit to the experimental data above 30 K. (inset: magnetization isotherms measured at 2 K with the magnetic field applied parallel to the $ab$ plane (solid squares) and the $c$ axis (open squares)). (e) Temperature dependence of electrical resistivity (inset: a magnified view at low temperature range showing AFM to PM transition). (f) Magnetoresistance (MR) measured in different temperatures from 2~K to 300~K.}
    \label{fig1}
\end{figure*}
\\
\section{Methods}
High quality single crystals of TbSbTe were grown using the self-flux method. The reactants were mixed in a stoichiometric composition and placed in an alumina crucible and then sealed in a quartz ampule in vacuum. The ampule was then placed in a muffle furnace and heated to $1000\degree$C at a rate of $10\degree$C/h. The temperature was held there for $10$~h and then cooled to $600
\degree$C at a rate of $2\degree$C/h, then excess Sb flux was separated using a centrifuge followed by a natural cooling of the sample. The chemical composition and phase purity of the crystals were ascertained through energy-dispersive X-ray (EDX) analysis, conducted using a FEI scanning electron microscope outfitted with an EDAX Genesis XM4 spectrometer. Furthermore, the crystal structure was validated by powder X-ray diffraction (XRD) analysis, performed on finely ground poly crystals of TbSbTe using a PANalytical X'pert Pro diffractometer equipped with $Cu-K_{\alpha}$ radiation. The orientation of the crystals selected for physical properties measurements was determined by means of back-scattered Laue diffraction, carried out on a Proto Manufacturing Laue-COS camera. The calculated crystal structure was visualized using the software VESTA version 3~\cite{vesta}, the 3D Brillouin zone (in the inset of Fig.~\ref{fig1}(b)) was generated using the software XCrySDen version 1.6.2~\cite{xcrysden}.

Magnetic measurements were carried out in the temperature interval 2-300~K in magnetic fields up to 14~T using a Quantum Design DynaCool-14 Physical Property Measurement System (PPMS). The heat capacity was measured from 1.72 to 210~K employing relaxation technique and two-$\tau$ model implemented in the PPMS platform.

Electrical transport measurements were performed in a thin bar-shaped piece of crystal (cross section: 244.5×155.5 µm$^2$, and distance between voltage probe: 644.5 µm) using a commercial PPMS platform. The magnetoresistance (MR) data have been symmetrized to eliminate possible contribution from the Hall resistance. The thermodynamic and electrical transport measurements were plotted using the OriginPRO (version 2016a) software package developed by Origin Lab.

The electronic band structure measurements were performed at High energy resolution spectrometer (HERS): Angle-resolved photoemission spectroscope (ARPES) at Advanced Light Source (ALS) beamline end station 10.0.1.1. The energy resolution was set to be better than 20~meV and the angular resolution was better than 0.2\degree~for all the measurements~\cite
{Neupane_2016}. The samples were mounted on copper sample holders, then posts were attached to their upper surface using silver epoxy. Then they were transferred to the high vacuum ARPES chamber and cleaved $in~situ$ at a pressure better than $10^{-10}$~torr and measurements were performed at a temperature of 18~K. All the ARPES data presented in the main manuscript and supplementray material were analyzed using the Igor pro 8.0.4 software package developed by WaveMetrics.

The {\it ab initio} calculations based on density functional thoery (DFT) were performed using the projector augmented-wave (PAW) potentials~\cite{blochl.94} implemented in the Vienna Ab initio Simulation Package ({\sc Vasp}) code~\cite{kresse.hafner.94,kresse.furthmuller.96,kresse.joubert.99}. Calculations are made within the generalized gradient approximation (GGA) in the Perdew, Burke, and Ernzerhof (PBE) parameterization~\cite{perdew.burke.96}.
The energy cutoff for the plane-wave expansion was set to $300$~eV, while $f$ electrons were treated as a core states.
In the case of magnetic calculations (with $f$-electrons treated as valence states), we assumed an AFM configuration, and applied a Hubbard $U = 6$~eV within the Dudarev approach~\cite{dudarev1998electron}.
Optimizations of structural parameters (lattice constants and atomic positions) are performed in the primitive unit cell using the $15 \times 15 \times 7$
 ${\bm k}$--point grid in the Monkhorst--Pack scheme~\cite{monkhorst.pack.76}
As a break of the optimization loop, we take the condition with an energy difference of $10^{-6}$~eV and $10^{-8}$~eV for ionic and electronic degrees of freedom. The topological properties, as well as the electronic surface states, were studied using the tight binding model in the maximally localized Wannier orbitals basis~\cite{marzari.mostofi.12, marzari.vanderbilt.97, souza.marzari.01}. This model was constructed from exact DFT calculations in a primitive unit cell (containing one formula unit), with $10 \times 10 \times 6$ $\Gamma$-centered  ${\bm k}$--point grid, using the {\sc Wannier90} software~\cite{pizzi.vitale.20}.
During the calculations, the $f$ electrons of Tb were treated as a core state.
The electronic surface states were calculated using the surface Green's function technique for a semi-infinite system~\cite{sancho.sancho.85}, implemented in {\sc WannierTools}~\cite{wu.zhang.18}.
\begin{figure*}
	\centering
	\includegraphics[width=\linewidth]{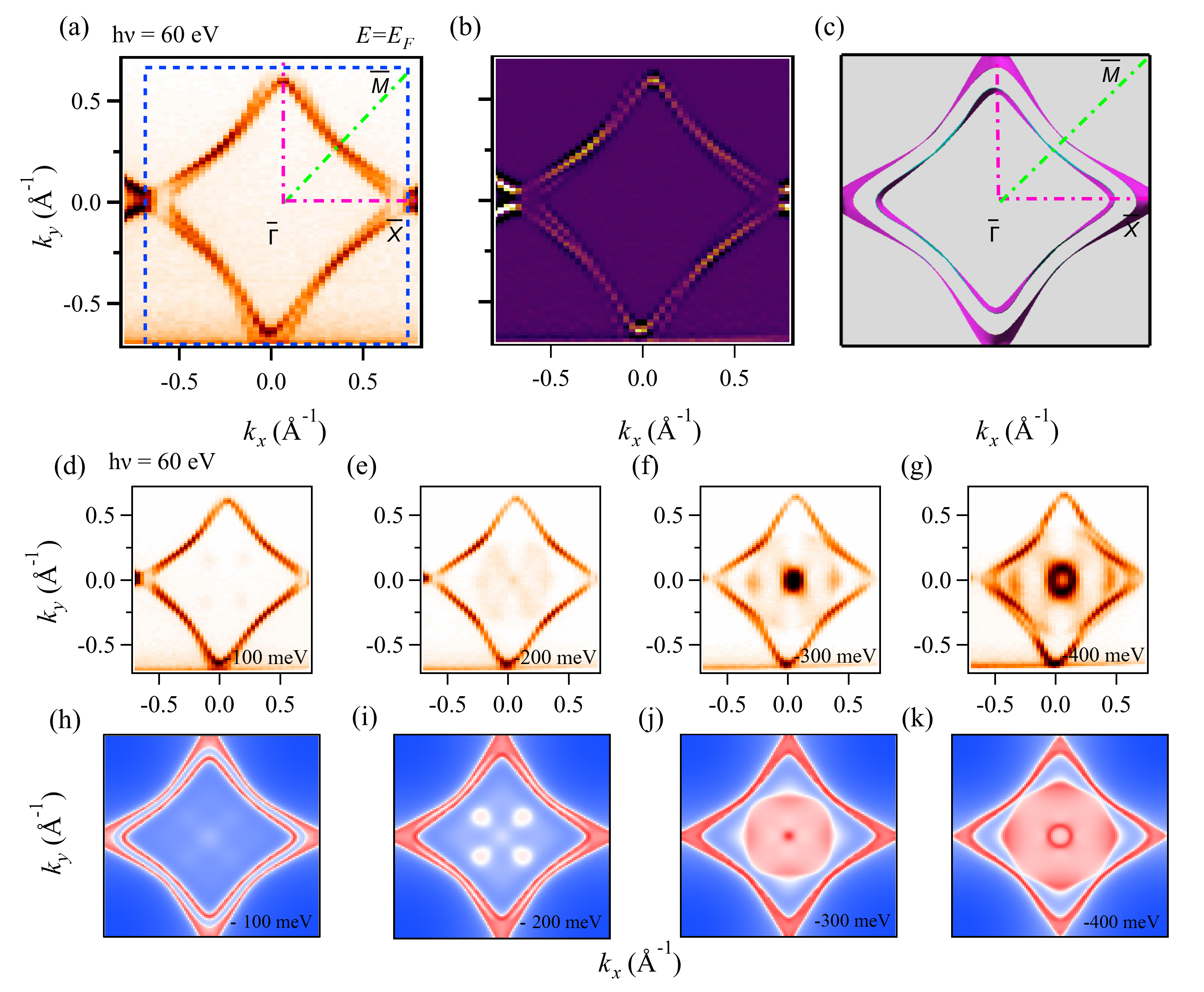}
	\caption{\justifying Fermi surface and Constant energy contour maps of TbSbTe. (a) Experimentally observed Fermi surface map of TbSbTe at an incident photon energy of 60~eV, the surface Brillouin zone is marked with a blue-dashed square and all the HS points ($\overline{\Gamma}$, $\overline{\text{M}}$ and $\overline{\text{X}}$) are marked. (b) Second derivative (SD) of the experimental Fermi surface. (c) Fermi surface obtained from DFT based calculations with the inclusion of SOC. (d-g) Experimentally observed constant energy contour maps at 100~meV, 200~meV, 300~meV and 400~meV below the Fermi surface, (h-k) theoretically calculated constant energy contours, with their binding energies marked on each of the contour plots. The ARPES measurements were done at Advanced Light Source (ALS) beamline 10.0.1.1 at a temperature of 18~K. }
 \label{fig2}
\end{figure*}\\

\section{Results and Discussion}
\textit{\textbf{Crystal and electronic structure.}}~Similar to other $Ln$SbTe materials, TbSbTe also crystallizes in the tetragonal $P4/nmm$ space group (No. 129) having the unit cell parameters $a=b=4.25$~\AA~and $c= 9.212$~\AA~\cite{Tb111}. 
The atoms are located in the high symmetry Wyckoff positions with Tb at $(2c)$ position: ($\frac{1}{4}$, $\frac{1}{4}$, $0.2759$), Sb at $(2a)$ position: ($\frac{3}{4}$, $\frac{1}{4}$, $0$) and Te at $(2c)$ position: ($\frac{1}{4}$, $\frac{1}{4}$, $0.6244$)~\cite{plokhikh_Ln, Tb111, Tb111_neutron} [Laue X-ray diffraction pattern and energy-dispersive X-ray spectroscopy (EDX) spectrum are presented in Supplementary Material (SM)~\cite{SM}, Fig.~S1]. The crystal structure of TbSbTe presented in Fig.~\ref{fig1}(a), comprises stacking of two Te-Tb-Te-Tb zigzag slabs having a $(\frac{1}{2}, \frac{1}{2},z)$ shift relative to each other that sandwich each square-planar Sb net when viewed along the $c$ axis~\cite{Tb111_neutron} [see Figs.~S2(a-c) in the SM~\cite{SM}].

The Brillouin zone (BZ), its 2D projection onto the (001) plane, and the positions of the nodal lines (along with the symmetries protecting them) are presented in SM Fig.~S2(d)~\cite{SM}. Fig.~\ref{fig1}(b) displays the calculated bulk band structure with (blue bands) and without (yellow bands) the effect of SOC, in PM phase. The locations corresponding to the nodal points observed in the experimental findings are indicated with green boxes. The nodal points can be tracked along the $k_z$ plane to the parallel HS directions to perceive their inclusion in a continuum of intersections. The bands of interest, taking part in formation of the nodal line features are marked with CB1, CB2, VB1 and VB2 in Fig.~\ref{fig1}(b). Schematics for better understanding the nodal features are added in SM Figs.~S2(e) and (f)~\cite{SM}. The bulk band structures in both the nonmagnetic and AFM phases, calculated with enhanced momentum resolution, are presented in Fig.~S3~\cite{SM}. The numerically obtained nodal lines, originating from these band crossings, are presented in Fig.~S4 of the SM~\cite{SM}. In addition to the symmetry-enforced nodal lines along the X-R and M-A directions, several accidental band crossings are also observed along the Brillouin zone boundaries.

\textit\textbf{{Bulk thermodynamic and electronic transport properties.}}~ 
The heat capacity data presented in Fig.~\ref{fig1}(c) corroborates the AFM order in TbSbTe. The AFM phase transition manifests itself as a distinct $\lambda$ type anomaly in $C_p$(T), the position of which perfectly coincides with $T_N$ determined from the magnetic data. Above about 100~K, $C_p$(T) saturates at a value of 75.7~J/mol K that is very close to the Dulong--Petit limit $3nR = 74.79$~J/mol$\cdot$K ($n = 3$ is the number of atoms per unit formula, and R stands for the gas constant). It should be noted that this finding notably differs from the results published before~\cite{Tb111}. Here, it should also be emphasized that the quantitative analysis of $C_p$(T) of TbSbTe, either within the framework of the simplified Debye model ($C\sim T^3$) or full Debye or Debye-Einstein formulas, is difficult without knowing the Schottky contribution from crystalline electric field excitations, which cannot be neglected at temperatures $T > 6$~K. In contrast, $C_p(T)$ in the AFM region is likely dominated by magnon contribution, which may be complex in nature, as suggested by the neutron diffraction results indicating the presence below $T_N$ of several competing commensurate and incommensurate magnetic orderings~\cite{Tb111_neutron}.

Fig.~\ref{fig1}(d) shows the temperature dependence of the magnetic susceptibility $(\chi \equiv M/H)$ of TbSbTe measured in a magnetic field applied along the crystallographic $c$-axis. In agreement with the previous reports~\cite{Tb111_neutron, Tb111}, $\chi$(T) shows a sharp maximum, marking the transition from PM to AFM state. The N\'eel temperature derived from this data is $5.8$~K, that is lower than $T_{N} = 6.4$~K given in Ref.~\cite{Tb111_neutron}, yet fairly close to $T_{N} = 6.0$~K determined in Ref.~\cite{Tb111}. As can be inferred from Fig.~\ref{fig1}(d), in the PM state, $\chi$(T) can be approximated by the modified Curie-Weiss (MCW) law, $\chi(T) = \chi_{o} + \frac{C}{(T -\theta_{p})}$, with the parameters $\chi_o = -2.4\times10^{-3}$~emu/mol, $C = 11.6$~emu K/mol and $\theta_{p} = -13.7$~K. The term $\chi_{o}$  is the sum of all the temperature-independent contributions to the magnetic susceptibility due to conduction electrons, core electrons, polarization effects, and crystal electric field effect. The effective magnetic moment $\mu_{eff}$, calculated from the value of the Curie constant C, equals  9.61~$\mu_B$/f.u, which is very close to the theoretical prediction for free Tb$^{3+}$ ion (9.72 $\mu_B$). The negative sign of the PM Curie temperature $\theta_{p}$ signals the predominance of AFM exchange interactions, in line with the AFM character of the electronic ground state in TbSbTe. It is worth noting that the values of $\mu_{eff}$ and $\theta_p$ is in good agreement with those reported in Ref.~\cite{Tb111}, $\mu_{eff}$ = 9.76 and $\theta_{p} = -12$~K for $H \parallel ab$. In Ref.~\cite{Tb111_neutron}, a polycrystal was measured so no direct comparison is justified in the case of a highly anisotropic material like TbSbTe.
\begin{figure*}
	\centering
	\includegraphics[width=\linewidth]{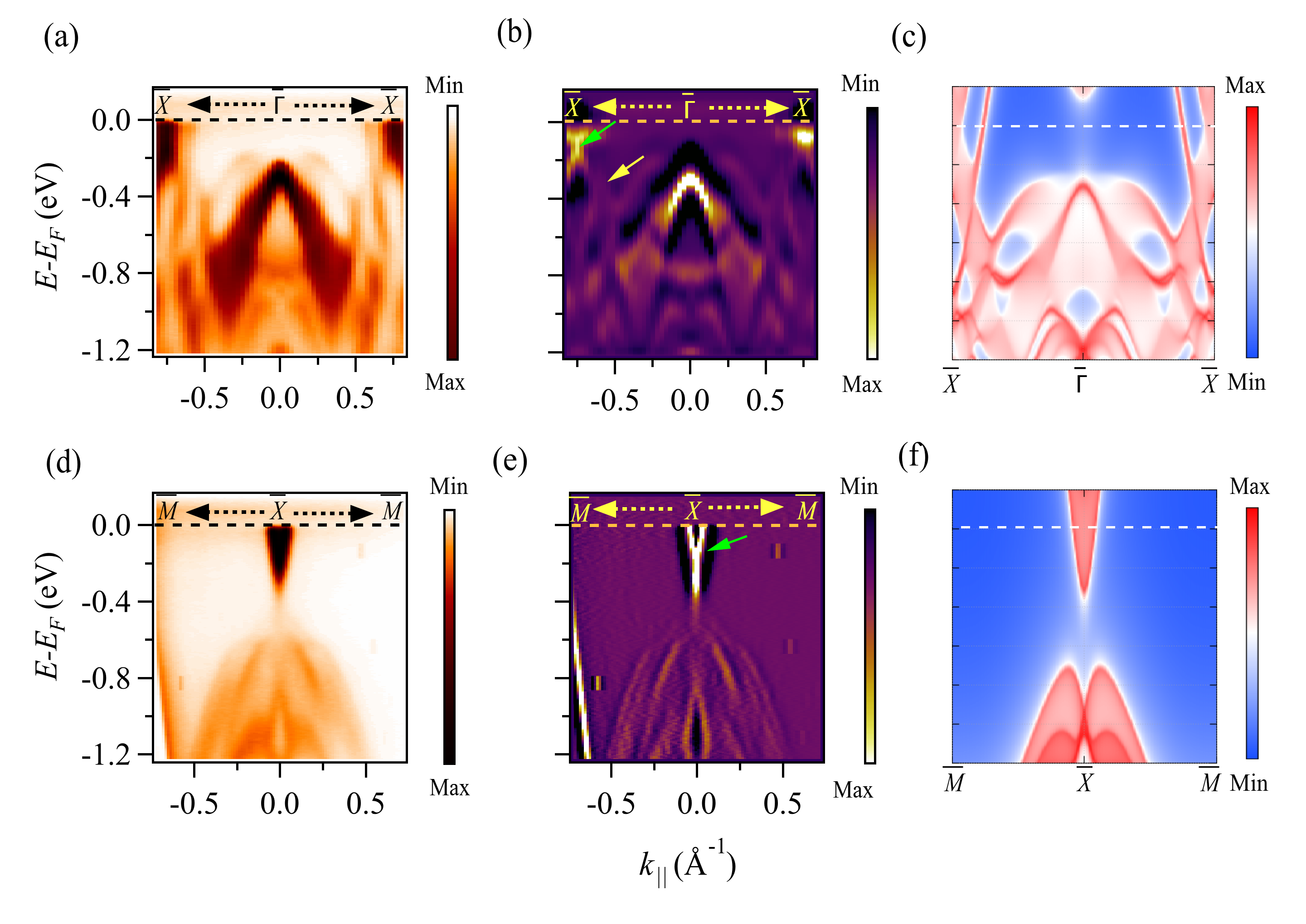}
	\caption{\justifying Observation of nodal line features parallel to the $\overline{\text{X}}$-$\overline{\Gamma}$-$\overline{\text{X}}$ and $\overline{\text{M}}$-$\overline{\text{X}}$-$\overline{\text{M}}$ directions. (a) Electronic dispersion map along $\overline{\text{X}}$-$\overline{\Gamma}$-$\overline{\text{X}}$ direction, (b) second derivative of the dispersion map along $\overline{\text{X}}$-$\overline{\Gamma}$-$\overline{\text{X}}$, the nodal points which are parts of nodal lines are marked with green (protected by nonsymmorphic symmetry) and yellow (not protected by nonsymmorphic symmetry) arrows, respectively. (c) the theoretically calculated surface projected band structure along $\overline{\text{X}}$-$\overline{\Gamma}$-$\overline{\text{X}}$. (d) dispersion maps along $\overline{\text{M}}$-$\overline{\text{X}}$-$\overline{\text{M}}$, (e) their second derivative and (f) theoretically calculated surface projected band structure. All the ARPES dispersion maps were collected at beam line 10.0.1.1 in Advanced Light Source (ALS) at a temperature of 18~K.}
  \label{fig3}
\end{figure*}

Inset of Fig.~\ref{fig1}(d) shows the magnetization $M(H)$ in TbSbTe measured at $T$ = 2~K in external magnetic field parallel to c axis (hard axis) and the ab crystallographic basal plane (easy axis). In agreement with the results shown in Ref.~\cite{Tb111}, the isotherm (with $\mu_{0}H\parallel ab$) exhibits two inflection points of metamagnetic-like character.
The occurrence of multi step changes in $M(H)$ can be associated with a complex non-collinear magnetic structure in zero magnetic field that at $T = 2$~K is defined by two propagation vectors $\mathbf{k}_1$ = ($\frac{1}{2}$ 0 0) and $\mathbf{k}_2$ = ($\frac{1}{2}$ 0 $\frac{1}{4}$
)~\cite{Tb111_neutron}. It is known that multi-k spin structures are susceptible to external magnetic field which often results in multiple field-induced rearrangements.

The temperature dependence of the electrical resistivity $\rho$(T) measured within the tetragonal plane of TbSbTe is shown in Fig.~\ref{fig1}(e). Both the magnitude and the shape of $\rho$(T) reflect a semimetallic character of the compound. The $\rho$(T) curve features a broad hump around 150~K, which may be associated to the interplay of gradual damping in charge carriers concentration and increase in scattering cross-section both occurring with decreasing temperature~\cite{Tb111, Tb111_neutron, Gebauer, HoSbTe_transport}. Another feature in $\rho$(T) is an upturn in the vicinity of $T_N$ followed by a sharp peak near 5~K (see the inset to Fig.~\ref{fig1}(e)) that manifests the onset of the ordered state. The prior study revealed a broad hump in $\rho$(T) occurring around 240~K , however no clear anomaly at T$_N$ was detected. The difference in the hump position may be attributed to diverse carrier concentrations in the crystals investigated. It seems likely that the same factor is responsible for the dissimilarities observed near T$_N$.

The transverse magnetoresistance, (MR)$= [\rho(H)/\rho(0)-1]\times100$, of TbSbTe was measured with electric current flowing in the basal plane and magnetic field directed along the tetragonal c-axis at several temperatures below and above $T_N$ (see Fig.~\ref{fig1}(f)). At 2~K, MR reaches in 9~T a value of 20$\%$, which quantitatively agrees with the previous study~\cite{Tb111}.

\textit{\textbf{Observation of nodal line states.}}~
In order to reveal the electronic structure of the compound, the Fermi surface (FS) map along with its second derivative (SD), collected at 60~eV incident photon energy are presented in the top panel of Figs.~\ref{fig2}(a) and (b), respectively. As can be inferred from Fig.~\ref{fig2}(c), the experimental ARPES data can be very well reproduced in the calculations based on the density functional theory (DFT) (presented in Figs.~\ref{fig2}(h-k)). At the $k_{x}-k_{y}$ plane, the FS consists of two diamond shaped iso-centric (around $\Gamma$ point) sheets. Examination of the constant energy contours (CECs) at higher binding energies (see Figs.~\ref{fig2}(d-g)) below the Fermi level indicates convergence of the two-sheeted diamond-like structure into a singular sheet. This coalesced single sheet retains the single sheet character till 300~meV of binding energy. However, near 500$\sim$600~meV (see CECs measured at higher binding energy presented in SM Fig.~S5, FS and CECs measured at different incident photon energies are also presented in SM Fig.~S6), the layers gradually diverge and reverts to its dual layer nature (for detailed analysis see SM Fig.~S7)~\cite{SM}. 

\begin{figure*}[htb!]
	\centering
	\includegraphics[width=\linewidth]{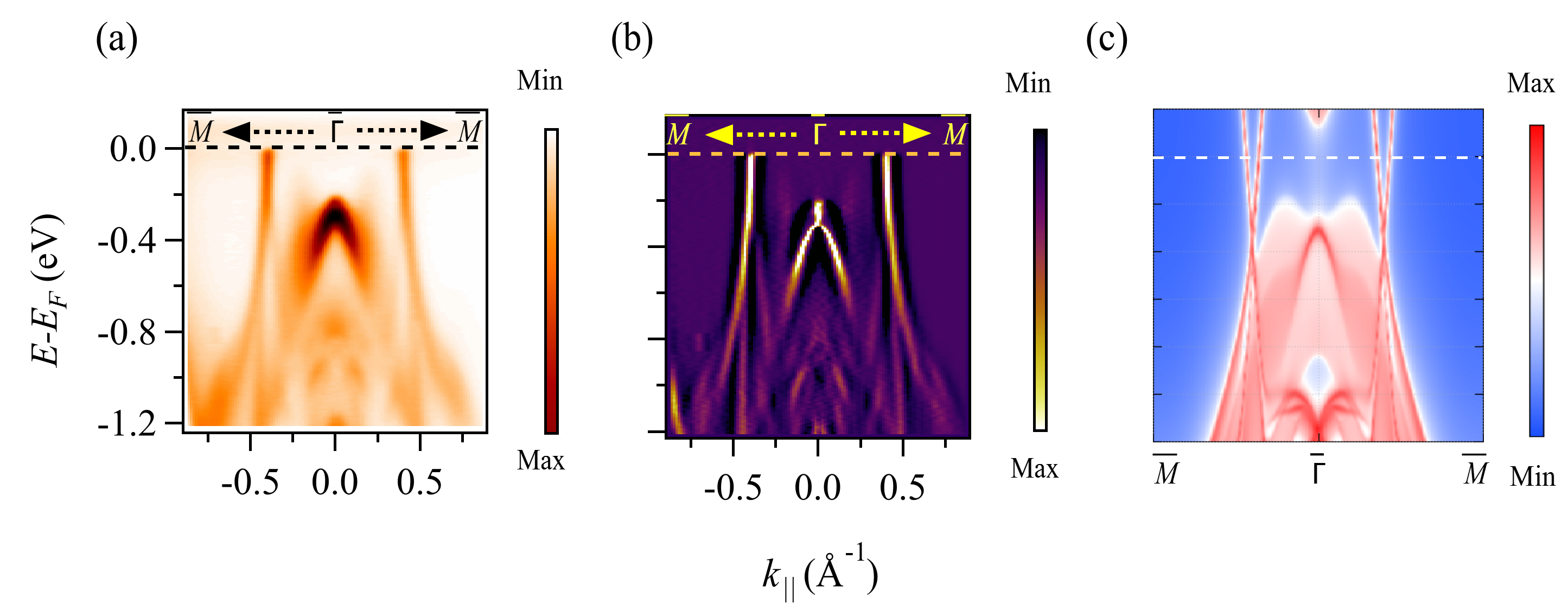}
	\caption{\justifying Observation of nodal states at the $\overline{\text{M}}-\overline{\Gamma}-\overline{\text{M}}$ direction.~(a) Dispersion map along $\overline{M}-\overline{\Gamma}-\overline{\text{M}}$ HS direction measured with 60~eV incident photon energy at a temperature of 18~K at ALS beamline 10.0.1.1. (b) Second derivative of $\overline{\text{M}}-\overline{\Gamma}-\overline{\text{M}}$ dispersion map, and (c) theoretically calculated surface band map along $\overline{\text{M}}-\overline{\Gamma}-\overline{\text{M}}$.}
    \label{fig_04}
\end{figure*}
To better understand the band crossings in the band structure of TbSbTe, we investigated the ARPES dispersion maps along different HS directions ($\overline{\Gamma}$-$\overline{\text{X}}$, $\overline{
\Gamma}$-$
\overline{\text{M}}$ and $\overline{\text{X}}$-$\overline{\text{M}}$). The dispersion maps along $\overline{\Gamma}$-$\overline{\text{X}}$ and $\overline{X}-\overline{M}$, measured at an incident photon energy of 60~eV, are presented in Figs.~\ref{fig3}(a) and~\ref{fig3}(d) (dispersion maps at different HS directions measured at other incident photon energies are presented in SM Fig.~S8~\cite{SM}). In Fig.~\ref{fig3}(a), along $\overline{\text{X}}$-$\overline{\Gamma}$-$\overline{\text{X}}$, two distinct crossing points can be observed (marked with green (protected by nonsymmorphic symmetry) and yellow (not protected by nonsymmorphic symmetry) arrows in~\ref{fig3}(b)), which are parts of the nodal lines, a common feature for the isostructural $Ln$SbTe materials~\cite{SmSbTe_sabin, La,CeSbTe_Lv}. These features were also reported in recent theoretical studies on TbSbTe as well~\cite{Tb111}. The first crossing occurs at the $\overline{\text{X}}$ point which can be better observed in the SD plot of the dispersion map presented in Fig.~\ref{fig3}(b) (also see Fig.~S9 in the SM, which presents the dispersion map, its second derivative, and the momentum dispersion curve at an incident photon energy of 50~eV to ascertain the presence of the Dirac like crossing~\cite{SM}). 
The band crossing at $\text{X}$ point remains robust against consideration of SOC as its protected by the additional nonsymmorphic symmetry, a well-known phenomenon in this system of materials. The experimentally obtained dispersion maps at different incident photon energies, parallel to $\Gamma$-$\text{X}$ (or Z-R) direction are presented in SM Fig.~S8(a-e), which show the consistent presence of this crossing at the $\overline{\text{X}}$ point at all photon energies, suggesting that the crossing is a part of out of plane nodal line along $k_z$~\cite{SM}. To better visualize the crossings, second derivative plots of the dispersion cuts measured with various incident photon energies are presented in~S10~\cite{SM}. The second crossing along the $\overline{\Gamma}$-$\overline{\text{X}}$ direction occurs approximately near 330~meV below the Fermi level. The positions of these Dirac crossings are plotted as a function of incident photon energies in SM Fig.~S11~\cite{SM}.Theoretical calculations suggest the existence of a small gap in the $k_z=0$ plane, which widens as we move toward the $k_z=\pi$ plane. However, this gap remained indiscernible in any of the dispersion cuts measured with different incident photon energies. 

Detailed analysis to examine the existence of these gaps are presented in SM Figs.~S12~\cite{SM}, exhibiting no signature of discontinuity in the intensity profiles, hinting absence of detectable gaps.

Next, we move onto the $\overline{\text{M}}$-$\overline{\text{X}}$-$\overline{\text{M}}$ HS direction. Fig.~\ref{fig3}(d), presents the dispersion cut along $\overline{\text{M}}$-$\overline{\text{X}}$-$\overline{\text{M}}$, its second derivative and its theoretically calculated surface spectrum (Figs.~\ref{fig3}(e) and (f)). 
An electron like pocket is present along this direction (dispersion map along $\overline{\text{M}}$-$\overline{\text{X}}$-$\overline{\text{M}}$ measured at other photon energies are presented in the SM Figs.~S7(k-o)~\cite{SM}), which is characteristic of $Ln$SbTe type materials~\cite{GdSbTe_hosen, Nd111, PrSbTe}. Bulk band calculations and surface projected spectrum, both exhibit similar pocket like feature suggesting presence of surface and bulk bands forming the pocket. 
Proceeding to the dispersion cut along the $\overline{\text{M}}$-$\overline{\Gamma}$-$\overline{\text{M}}$, presented in Fig.~\ref{fig_04}(a), two linear bands with different velocities converge to form an intersection, which is well reproduced in surface projected calculations presented in Fig.~\ref{fig_04}(c). Again, contrasting the experimental observations, bulk calculation (Fig.~\ref{fig1}(b)) suggests that, a gap occurs at the intersection of bands CB2 and VB2 (see Fig.~\ref{fig1}(b)), which persists in a reduced form, if SOC is taken into account. This gap was also beyond the detection limits of the ARPES detector. Another notable aspect is a hole pocket like feature which can also be seen centering the $\Gamma$ (or $Z$) point in cuts parallel to both $\Gamma-\text{X}$ and $\Gamma-\text{M}$ directions (see Figs.~\ref{fig3} and~\ref{fig_04} respectively), throughout all the incident photon energies (presented in SM Fig.~S7(a-e) and (f-j) respectively~\cite{SM}), but its spectral intensity was highly sensitive to the incident photon energy.\\

\begin{figure*}[htb!]
	\centering
	\includegraphics[width=\linewidth]{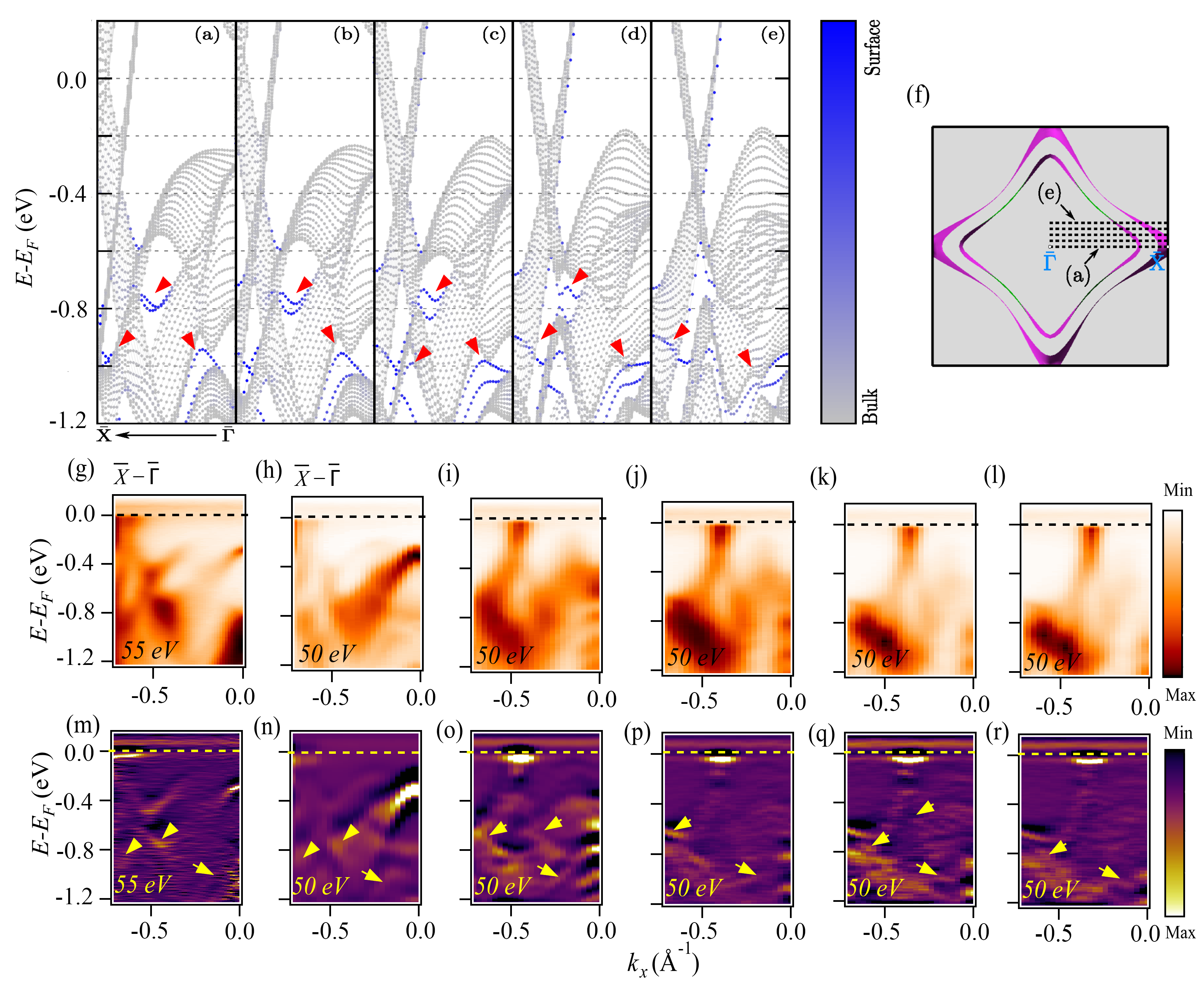}
	\caption{\justifying Surface states parallel to $\overline{\Gamma}$-$\overline{\text{X}}$ direction.~Panels (a-e) show the calculated band structures for the slab geometry along cuts presented in panel (f). The surface originated bands (blue) are marked with red arrows. ARPES experimental spectra along $\overline{\Gamma}$-$\overline{\text{X}}$ HS direction recorded with (g) 55~eV and (h) 50~eV incident photon energies, and cuts at the corresponding positions of panels (b-e) are presented in the panels (i-l). To resolve the band features better, the second derivatives of (g-l) are presented in (m-r), here the surface originated bands are marked with yellow arrows. The experimental measurements were performed at a temperature of 18~K at ALS beamline 10.0.1.1.}
    \label{fig_05}
\end{figure*}

\textit{\textbf{Momentum dependent surface states parallel to \texorpdfstring{$\overline{\Gamma}$-$\overline{\text{X}}$}{Gamma-X}.}}~
A detailed analysis of the surface states based on theoretical and experimental results are presented in Fig.~\ref{fig_05}. The theoretically calculated surface Green function and slab band structures are presented in Fig. panels~\ref{fig_05}(a-e), along and parallel to the $\overline{\Gamma}$-$\overline{\text{X}}$ HS direction (see the cuts in the Fermi surface map in Fig.~\ref{fig_05}(f)). The grey bands corresponds to bulk and blue bands corresponds to surface originated bands.
The band features centering the $\Gamma$ and X HS points are predominately bulk originated. The theoretical calculations are well reproduced in the experimental ARPES based observations presented in Figs.~\ref{fig_05}(g-l) and their corresponding second derivatives in Figs.~\ref{fig_05}(m-r). Figs.~\ref{fig_05}(g) and h are ARPES dispersion cuts measured with 55 and 50~eV of incident photon energies along $\overline{\Gamma}$-$\overline{\text{X}}$, exhibiting surface bands forming pockets around the $\overline{\Gamma}$ HS point. From $E_F$ to 600~meV binding energy, a linearly dispersing surface band is observed within the bulk bands. Below that, in the proximity of 800~meV binding energy, two branches of surface bands are observed. Another set of surface bands are observed right at the $\overline{\text{X}}$ HS point. In total, we see four regions containing the surface bands. Cuts parallel to $\overline{\Gamma}$-$\overline{\text{X}}$ directions, presented in Figs.~\ref{fig_05}(h-l) and their corresponding second derivative plots in~\ref{fig_05}(n-r), exhibit strong momentum dependence of the surface bands.\\\\
\section{Discussion}
Thermodynamic, magnetic and transport measurements on TbSbTe single crystals coherently exhibit the AFM to PM phase transition near 5.8~K. The electronic band structure of TbSbTe, in ARPES spectrographs, exhibit two isocentric diamond shaped sheets around the $\Gamma$ point similar to other $Ln$SbTe and ZrSiS type materials. For TbSbTe the double-sheet is consistently present throughout the FS, unlike heavier members of $Ln$SbTe (for an example, HoSbTe~\cite{HoSbTe_transport}), where the bands tend to gap out while radially going towards the $\overline{\Gamma}$-$\overline{\text{M}}$ direction. Another feature of this FS is that, the bands are not as well separated as in NdSbTe~\cite{Nd111} or PrSbTe~\cite{prsbte_yuan, PrSbTe}, nor coalesced into a single sheet like GdSbTe~\cite{GdSbTe_hosen}. 
The features of the nodal line observed in the cuts parallel to the $\overline{\Gamma}$-$\overline{\text{X}}$ are well reproduced in bulk electronic calculations. In space group $P4/nmm$, symmetries in combination with nonsymmorphic symmetry plays a crucial role in realizing enforced degeneracy in the vicinity of Fermi energy. This space group hosts a glide plane $\mathcal{\Tilde{M}}_z=\{\mathcal{M}_z|\frac{1}{2}\frac{1}{2}0\}$, screw axes, $\Tilde{C}_{2\nu,~(\nu= x,y)}=\{\mathcal{
C}_{2x}|\frac{1}{2}00\}$ and $\{\mathcal{\Tilde{C}}_{2y}|0\frac{1}{2}0\}$, mirror symmetries $\mathcal{\Tilde{M}}_x=\{\mathcal{M}_x|\frac{1}{2}00\}$, $\mathcal{\Tilde{M}}_{y}=\{\mathcal{M}_{y}|0\frac{1}{2}0\}$ and $\mathcal{\Tilde{M}}_{xy}=\{\mathcal{M}_{xy}|\frac{1}{2}\frac{1}{2}0\}$, and the spatial inversion symmetry $\mathcal{P}=\{\mathcal{P}|000\}$.
Dirac crossings observed in the HS directions parallel to $\Gamma$-X (with different incident photon energies, i.e. at different $k_z$ locations), right at the $\overline{\text{X}}$ points, are degeneracies enforced by the combined effect of glide plane, $\mathcal{\Tilde{M}}_z$ and time reversal symmetry, $\mathcal{T}$. These crossings are part of the nodal line that extends along the X-R HS direction. The combination of these symmetries ensures the robustness of this nodal line even under consideration of the effect of SOC. The screw axis symmetries, $\Tilde{C}_{2\nu}$ in conjugation with the inversion symmetry, $\mathcal{P}$ allows the other crossing (at the relatively higher energies), which also extends longitudinally along the $k_z$ axis. Mirror symmetry, $\mathcal{\Tilde{M}}_{xy}$ allows the crossings in the $\Gamma$-M HS direction, which are parts of the nodal line extending parallel to the M-A direction, often observed in the ZrSiS type systems. However, nodes in this direction were not observed conclusively in this study. Although, except for the nonsymmorphic symmetry enforced nodal line (along X-R, at the X point), consideration of SOC destabilizes the other nodal crossings. Despite the theoretical predictions of destabilized nodal line features, in experimental observations, the nodal lines remain ungapped. This phenomenon could potentially be ascribed to constrained $k_z$ resolution within the vacuum ultraviolet region, resulting in the energy cuts encompassing a spectrum of $k_z$ values rather than being confined to a specific $k_z$ value. An alternate explanation might stem from the prevalent incongruity between experimental results and theoretical predictions often observed in 4$f$ systems. Moreover, while the previously discussed symmetries preserve the degeneracies in the PM phase, consideration of magnetic ordering (i.e. AFM or FM magnetic ordering) usually leads to breaking of these degeneracies. In the electrical and thermodynamic measurements, TbSbTe undergoes a PM to AFM phase transition, thus we restrict our considerations to AFM ordering. AFM spin orientation of Tb along (001) direction elongates the $c$ lattice parameter into $2c$ (technically space group $P4/ncc$ with $\mathcal{T}$). This still guarantees the nodal line along the X-R direction. A similar analysis was also performed in Ref.~\cite{Schoop}.\\
\section{Conclusion}

To summarize, we present temperature and magnetic field-dependent thermodynamic and electric transport investigations, which consistently indicates an AFM to PM transition at $T_N = 5.8$~K. Concurrently, we studied the electronic band structure of a new member of the $Ln$SbTe family, TbSbTe, using high-resolution ARPES combined with theoretical first-principles calculations. Our electronic structure study revealed multiple band intersections, including a Dirac crossing, formed by bulk bands, in the $\Gamma$-X HS direction, which is part of a nodal line extending along X-R bulk direction. This crossing remains ungapped over $k_z$, as established in both our theoretical work considering the effect of SOC and our experimental ARPES-based observations. Another Dirac like crossing in this direction, was gapped in the theoretical calculations. These results implicate TbSbTe as a promising candidate for examining the interaction of symmetry, topology and SOC in $Ln$SbTe series of materials. \\

\vspace{2ex}

\section*{acknowledgments} 
\vspace{-0.1 cm}
M. N. acknowledges support from the National Science Foundation under the CAREER award DMR-1847962 and the Air Force Office of Scientific Research for MURI Grant No. FA9550-20-1-0322. N. V. acknowledges the support from the MPS Alliances for Graduate Education and the Professoriate--Graduate Research Supplements (AGEP--GRS).  D. K. and T.R. were supported by the National Science Centre (Poland) under research grant 2021/41/B/ST3/01141. V. B. and S. R. acknowledge the support from Idaho National Laboratory’s Laboratory Directed Research and Development (LDRD) program under the Department of Energy (DOE) Idaho Operations Office Contract DEAC07-05ID14517. K. G. acknowledges support from the Division of Materials Science and Engineering, Office of Basic Energy Sciences, Office of Science of the U. S. Department of Energy (U.S. DOE). A. P. acknowledges the support by National Science Centre (NCN, Poland) under Projects No.~2021/43/B/ST3/02166.
This research utilized resources of the ALS at the Lawrence Berkeley National Laboratory, which is a DOE Office of Science User Facility under Contract No. DE-AC02-05CH11231. Magnetic measurements were performed at the State University of New York (SUNY), Buffalo State, and supported by the National Science Foundation, Launching Early-Career Academic Pathways in the Mathematical and Physical Sciences (LEAPS-MPS
) program under Award No. DMR-2213412. We express our gratitude to Dr. Sung-Kwan Mo for providing valuable support with the beamline at the Advanced Light Source (ALS). \\

\raggedbottom
\clearpage
\pagebreak{}

\clearpage
\widetext
\begin{center}
\textbf{\large Supplemental Materials for: \\~\\\Large Electronic Structure of a Nodal Line Semimetal Candidate TbSbTe}\\
\end{center}
\setcounter{equation}{0}
\setcounter{figure}{0}
\setcounter{table}{0}
\setcounter{page}{1}
\makeatletter
\renewcommand{\theequation}{S\arabic{equation}}
\renewcommand{\figurename}{{Fig.}}
\renewcommand{\thefigure}{{{S\arabic{figure}}}}
\renewcommand{\bibnumfmt}[1]{[#1]}
\renewcommand{\citenumfont}[1]{#1}
\renewcommand{\tablename}{Supplementary Table}
\renewcommand{\thetable}{\arabic{table}}
\def\bibsection{\refname}
\renewcommand{\refname}{\noindent\textbf{Supplementary References}\\}

\begin{center}
\vspace{1 cm}
\textbf{SUPPLEMENTARY ARTICLES}   
\end{center}
\section*{S1. Single crystal characterization and theoretical band structure calculations}
Fig.~\ref{Tb_sm_01}(a) presents a photograph of single crystals of TbSbTe grown for the purpose of the present study. They were found by XRD to have the expected tetragonal crystal structure with the lattice parameters close to those reported in the literature~\cite{Gao2022, Plokhikh2022}. The natural phase of the as-grown crystals is (001), as proven by the Laue technique (see Fig.~\ref{Tb_sm_01}(b)). The XRD and EDX data indicated that the crystals are single-phase and homogeneous with the chemical composition very close to the ideal one (viz.~\ref{Tb_sm_01}(c)). Figs.~\ref{SM_fig_01}(a-c) visualizes the crystallographic unit cell of TbSbTe that can be represented as a stacking of Tb-Te stabs and quadratic Sb layers. The bulk Brillouin zone, its projection on the (001) plane and the positions of the nodal lines (along with the symmetries protecting/enforcing them) are shown in Fig.~\ref{SM_fig_01}(d). The nodal lines along the X--R high symmetry direction, are nonsymmorphic symmetry enforced crossings [glide plane symmetry, $\mathcal{\Tilde{M}}_z=\{\mathcal{M
}_z|\frac{1}{2}\frac{1}{2}0\}$ in combination with time reversal symmetry, $\mathcal{T}$], which are represented by the green broken lines in Figs.~\ref{SM_fig_01}(d). The allowed nodal line along X--R [screw axis, $\Tilde{C}_{2\nu,~(\nu= x,y)}=\{\mathcal{
C}_{2x}|\frac{1}{2}00\}$ and inversion symmetry, $\mathcal{P}$] is represented with yellow broken lines. The corresponding Dirac nodes are indicated with green and yellow dots in the (001) surface Brillouin zone~\cite{Topp2017}. 
To better understand the positions of the nodal lines, visualizations are presented in Figs.~\ref{SM_fig_01}(e) and (f).   
The calculated bulk band structure of TbSbTe, in the nonmagnetic and antiferromagnetic (AFM) phase is presented in the Fig.~\ref{bulk_band}(a) and (b) respectively. Plokhikh {\it et al}. reported the different magnetic structures of TbSbTe in details in Ref.~\cite{Plokhikh2023}. Such magnetic configurations are exclusively stabilized in large unit cells, where magnetic moments arrange into square blocks exhibiting AFM-like ordering. However, in the present case, these magnetic structures proved to be numerically unstable. The results presented in Fig.~\ref{bulk_band}(b) are based on DFT+U, where the effective correlation parameter (U), shifts the bands away from the Fermi energy, appearing as flat bands ranging from $\sim$-2~eV to $\sim$-4~eV and $\sim~$0.5~eV to $\sim$~2~eV. Thus, the magnetic schemes have minimal effect on the low lying electronic structure of the material. One significant change that occurs to the enforced band crossing at the X point at the AFM phase with SOC considered, is the crossing is gapped.  
The nodal lines obtained from different band crossings are visualized in Fig.~\ref{nodes}.

\section*{S2. ARPES based band structure measurements}
Angle resolved photoemission spectroscopy (ARPES) based measurements were carried out at Advanced Light Source (ALS) beamline endstation 10.0.1.1 at 18~K. The sample crystals were cleaved at ultra high vacuum better than $10^{-10}$~Torr. For the measurements of the ARPES maps, the energy resolution was set to be better than 20~meV and the angle resolution was better than 0.2\degree. The constant energy contours ranging from $-500$~meV to $-800$~meV, measured with 60~eV incident photon energy, are presented in Fig.~\ref{SM_fig_BZ}, which is a continuation of Fig.~2 of the main manuscript. The Fermi surfaces and the constant energy contours measured using 45~eV to 55~eV incident photon energies are displayed in SM Fig.~\ref{SM_fig_hv}. \\
\indent Fermi surface at all the photon energies persistently exhibit the double-sheet nature similar to the 60~eV data presented in the main manuscript. Fig.~\ref{SM13} presents a detailed analysis of ARPES data to show the dual sheet nature of the Fermi surface. Intensity profile was taken from two energy positions from the cut parallel to the $\overline{\text{M}}-\overline{\Gamma}$ direction (Fig.~\ref{SM13}(a)). The corresponding Fermi surface maps (cut C1) at two different energies (with h$\nu$ = 55 and 60~eV) are presented in Fig.~\ref{SM13}(b) and CEC corresponding to cut C2 is presented in Fig.~\ref{SM13}(c). Intensity profile taken from the broken red line in Fig.~\ref{SM13}(b), shown in Fig.~\ref{SM13}(d), exhibits two peaks, representing the two sheets, discussed above. \\
\indent Electronic dispersion maps measured with incident photon energies ranging from 45~eV to 65~eV are presented in SM Fig.~\ref{SM_fig_03}. SM Figs.~\ref{SM_fig_03}(a-e) show crossings in the dispersion cut parallel to $\overline{\Gamma}$-$\overline{\text{X}}$ direction which are part of nodal lines along $k_z$ direction at approximately $-150 \sim -200$~meV and $-330$~meV.
In SM Figs.~\ref{SM_fig_03}(f-j), the crossing near $-200$~meV is persistent along increasing incident photon energy along $\overline{\Gamma}-\overline{\text{M}}$. All of the aforementioned cuts host a strongly energy dependent hole pocket centered at $\Gamma$. In the energy range from the Fermi level down to $-0.4$~eV, SM Fig.~\ref{SM_fig_03}(k--o) reveals a pronounced energy-dependent Dirac node feature along the $\overline{\text{M}}$--$\overline{\Gamma}$ momentum cuts. The variation in intensity with incident photon energy indicates that these features arise from bulk electronic states. In contrast, the wing-like feature observed below $-0.5$~eV remains invariant with respect to photon energy, suggesting a surface-derived origin.\\
\indent SM Fig.~\ref{SM5} presents the ARPES dispersion map and second derivative of the dispersion cut at 50~eV, near $\overline{\Gamma}$-$\overline{\text{X}}$ direction. To better visualize the band features, the stacked momentum distribution curves (MDC) integrated within a span of 20~meV are presented in SM Fig.~\ref{SM5}(c). Second derivative plots of the dispersion cuts at different incident photon energies are presented in SM Fig.~\ref{SM8}, and the positions of the crossings are presented in SM Fig.~\ref{SM9}, as a function of incident photon energy. The crossings which are not protected by the nonsymmorphic symmetry were gapped in the theoretical bulk calculations; magnified views of those crossings are presented in Fig.~\ref{SM10}. The EDC taken at the corresponding position of the crossing ($\pm$0.01~\text{\AA}$^{-1}$), shown in the Fig.~\ref{SM10}(b) do not show any signature of gap opening, may be because of instrument resolution. The crossing appears to be slightly tilted, suggesting a type II nature. In Fig.~\ref{SM10}(c), we calculated the MDCs of Fig.~\ref{SM10}(a), and tracked the peak points connecting them (represented by blue broken lines).\\ \\

\clearpage
\begin{center}
\textbf{SUPPLEMENTARY FIGURES}   
\end{center}
\setcounter{figure}{0}
\begin{figure}[!hb]
	\centering
	\includegraphics[width=15 cm]{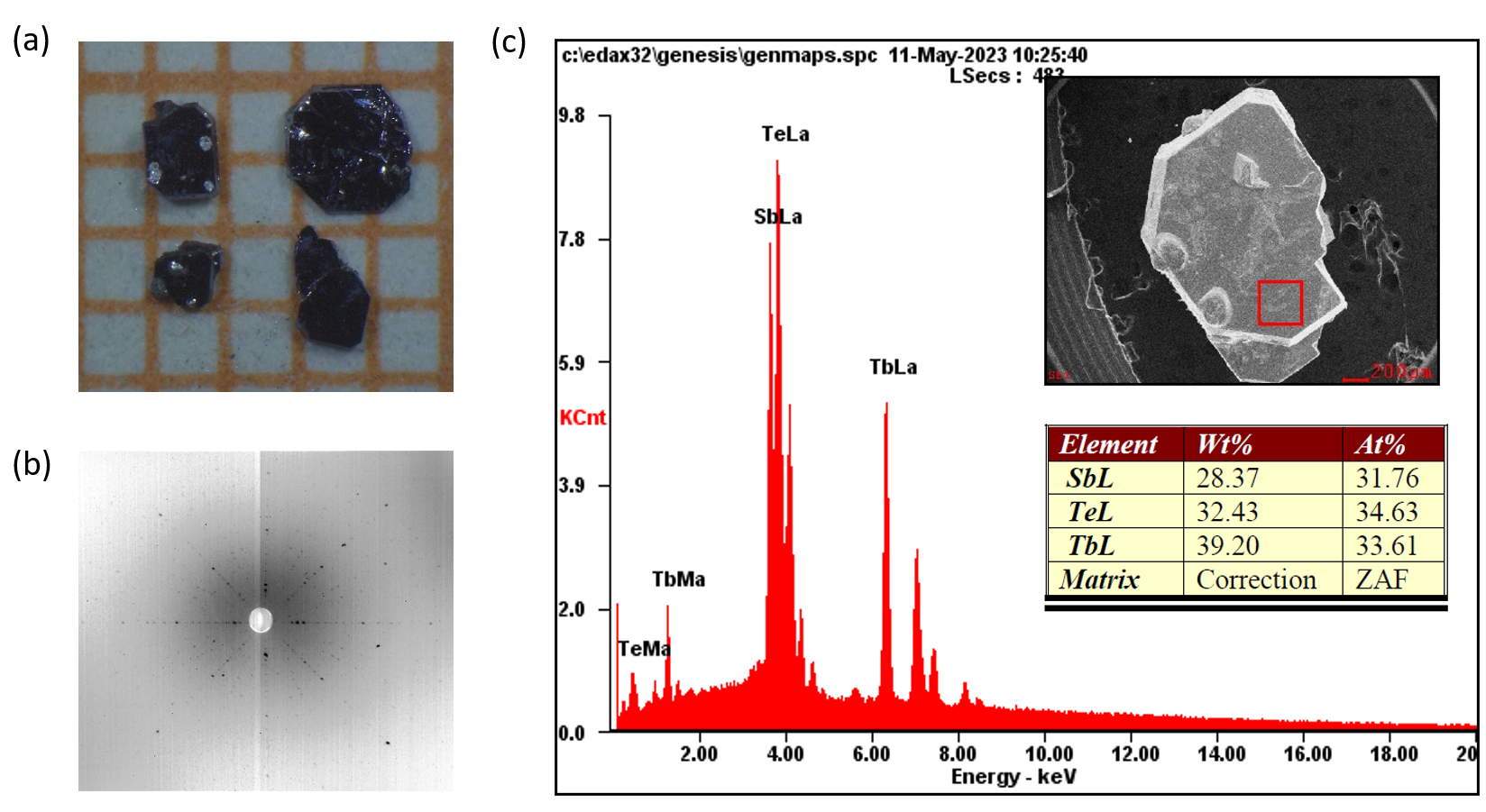}
	\caption{\justifying Crystal-chemical characterization of TbSbTe single crystals. (a) Photographs of single crystals selected for physical properties studies. (b) Laue X-ray diffraction pattern from natural (001) plane. (c) EDX spectrum and the chemical composition analysis of an exemplary area of crystalline specimen.}
  \label{Tb_sm_01}
\end{figure} 

\newpage

\begin{figure}[!hb]
	\centering
	\includegraphics[width=15 cm]{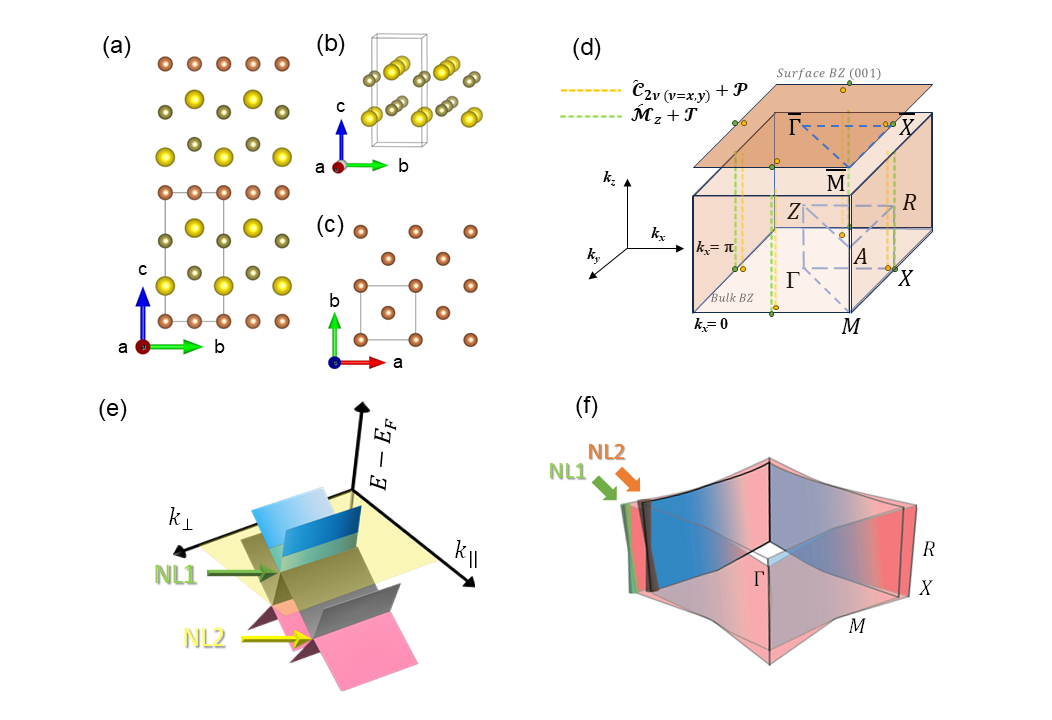}
	\caption{\justifying Crystal structure of TbSbTe. (a) Viewed along the a axis, (b) zigzag Tb (yellow)-Te(golden) layers, and (c) Sb (orange)-square planar layer. (d) Construction of a 3D bulk Brillouin zone (BZ) in momentum space and projection on its natural cleaving plane (001) surface. The positions/directions of the nodal points/nodal lines and the symmetries associated to them are represented by broken orange and green lines. (e) Schematics of the nodal lines in binding energy vs. k--plane and (f) in the 3D momentum space.}
  \label{SM_fig_01}
\end{figure}

\begin
{figure}[!htb]
	\centering
	\includegraphics[width=13 cm]{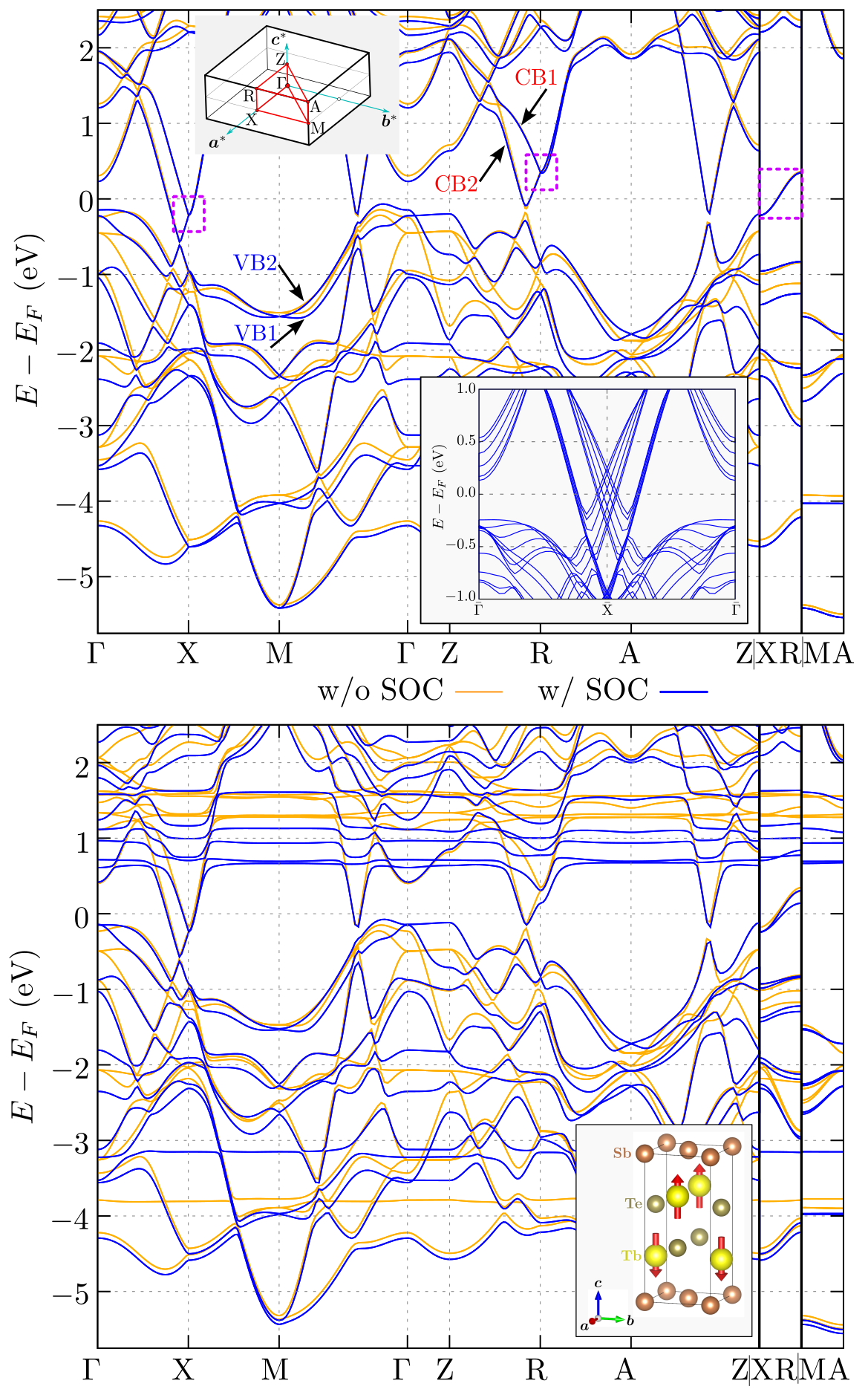}
	\caption{\justifying Calculated bulk band structure of TbSbTe in (a) nonmagnetic phase, inset: (top) 3D bulk Brillouin zone, (bottom) a magnified view of band crossings in the X-$\Gamma$-X high symmetry direction. (b) Bulk band structure in antiferromagnetic (AFM) phase (inset: magnetic structure of TbSbTe). }
  \label{bulk_band}
\end{figure}

\begin{figure}[!ht]
	\centering
	\includegraphics[width=0.8\linewidth]{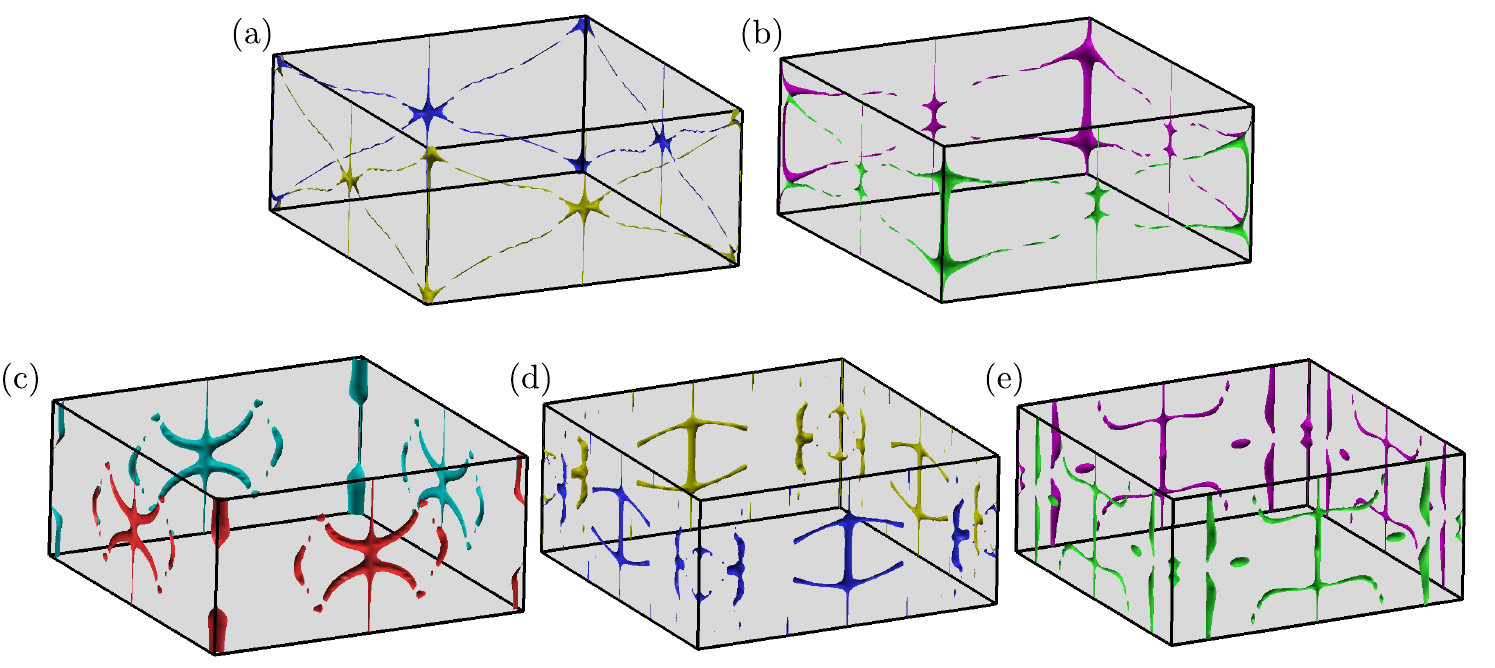}
	\caption{\justifying The nodal lines obtained from the band crossings in the nonmagnetic calculation presented in Fig. S3(a), the black frame corresponds to the 3D Brillouin zone.}
  \label{nodes}
\end{figure}
\begin{figure}[!ht]
	\centering
	\includegraphics[width=15 cm]{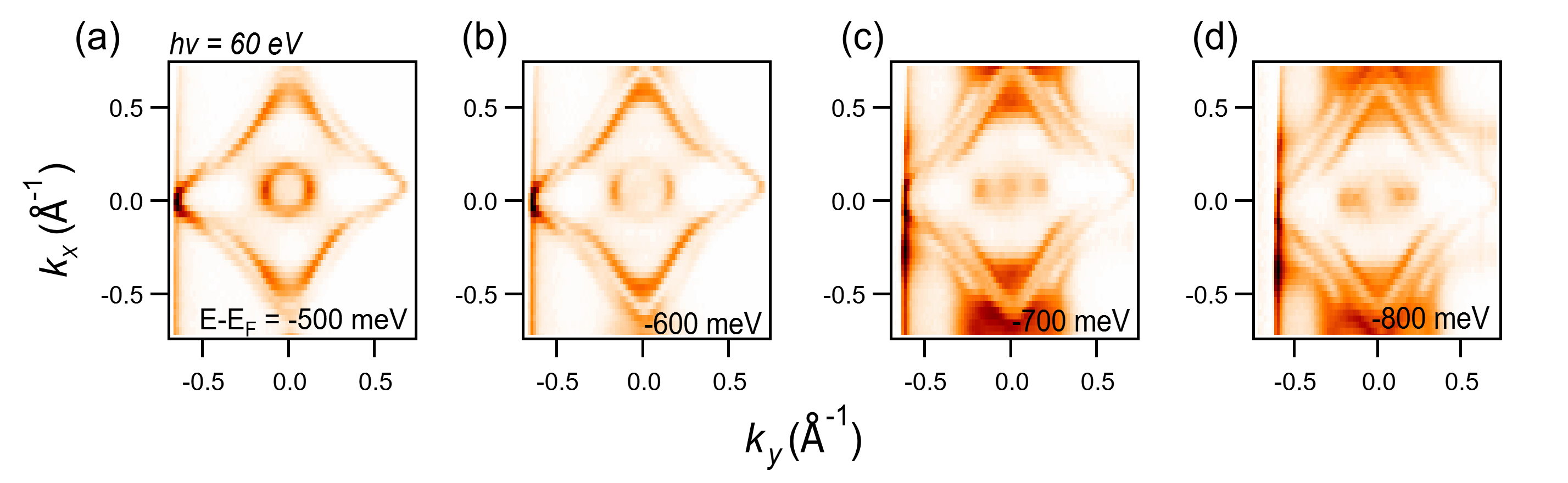}
	\caption
{\justifying Constant energy contours (CEC) from TbSbTe ARPES spectra measured with 60~eV incident photon energy, at several binding energies (indicated at the bottom of each subplots) collected at beamline 10.0.1.1 in Advanced Light Source (ALS) at a temperature of 18~K.}
  \label{SM_fig_BZ}
\end{figure}

\begin{figure}[!hb]
	\centering
	\includegraphics[width=15 cm]{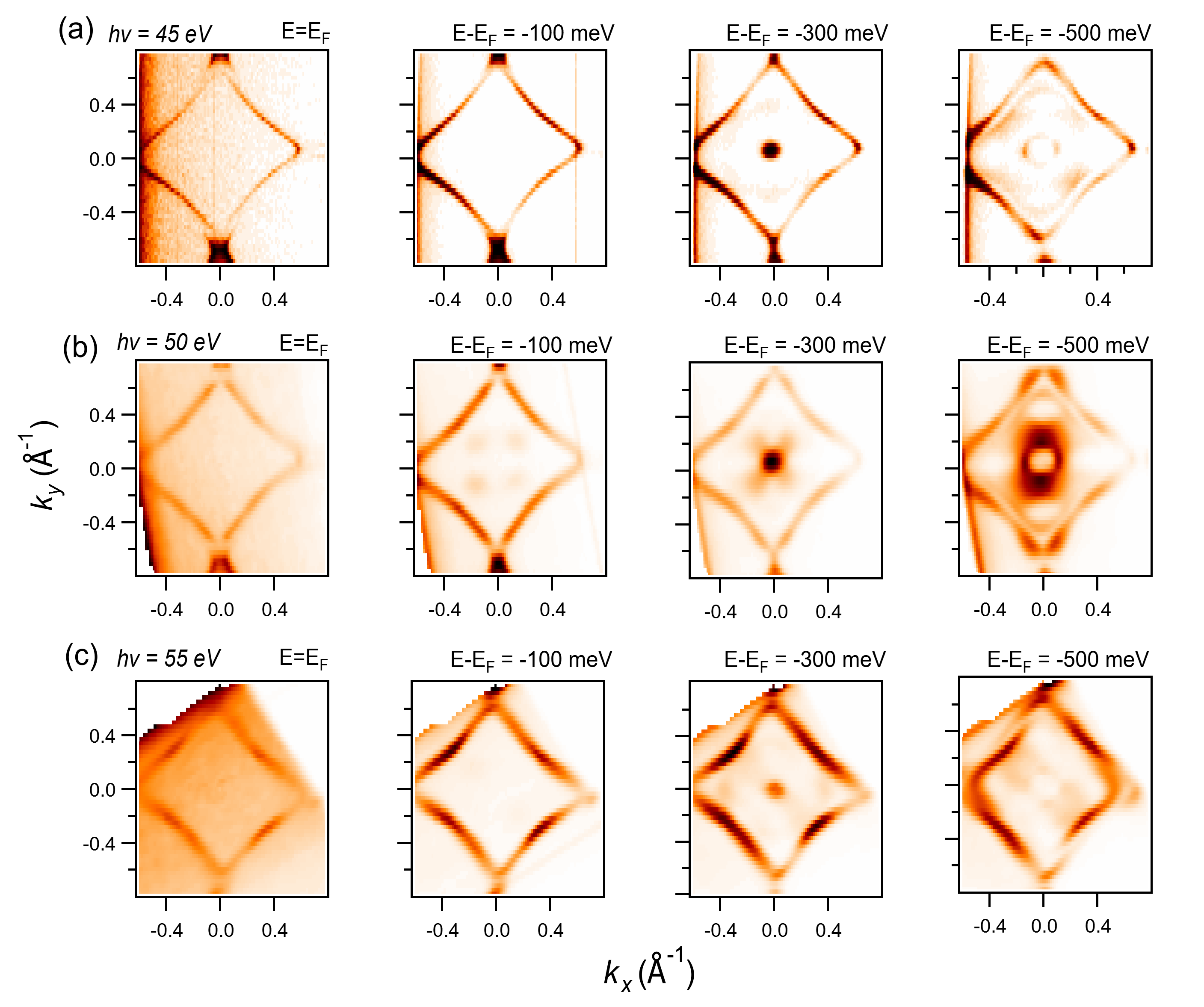}
   
	\caption
{\justifying Fermi surface and constant energy contours (CEC) of TbSbTe ARPES spectra at incident photon energies (a) 45~
eV, (b) 50~eV and (c) 55~eV. ARPES measurements were performed at ALS beamline 10.0.1.1 at a temperature of 18~K.}
  \label{SM_fig_hv}
\end{figure}

\begin{figure*}[!ht]
	\centering
	\includegraphics[width=13cm]{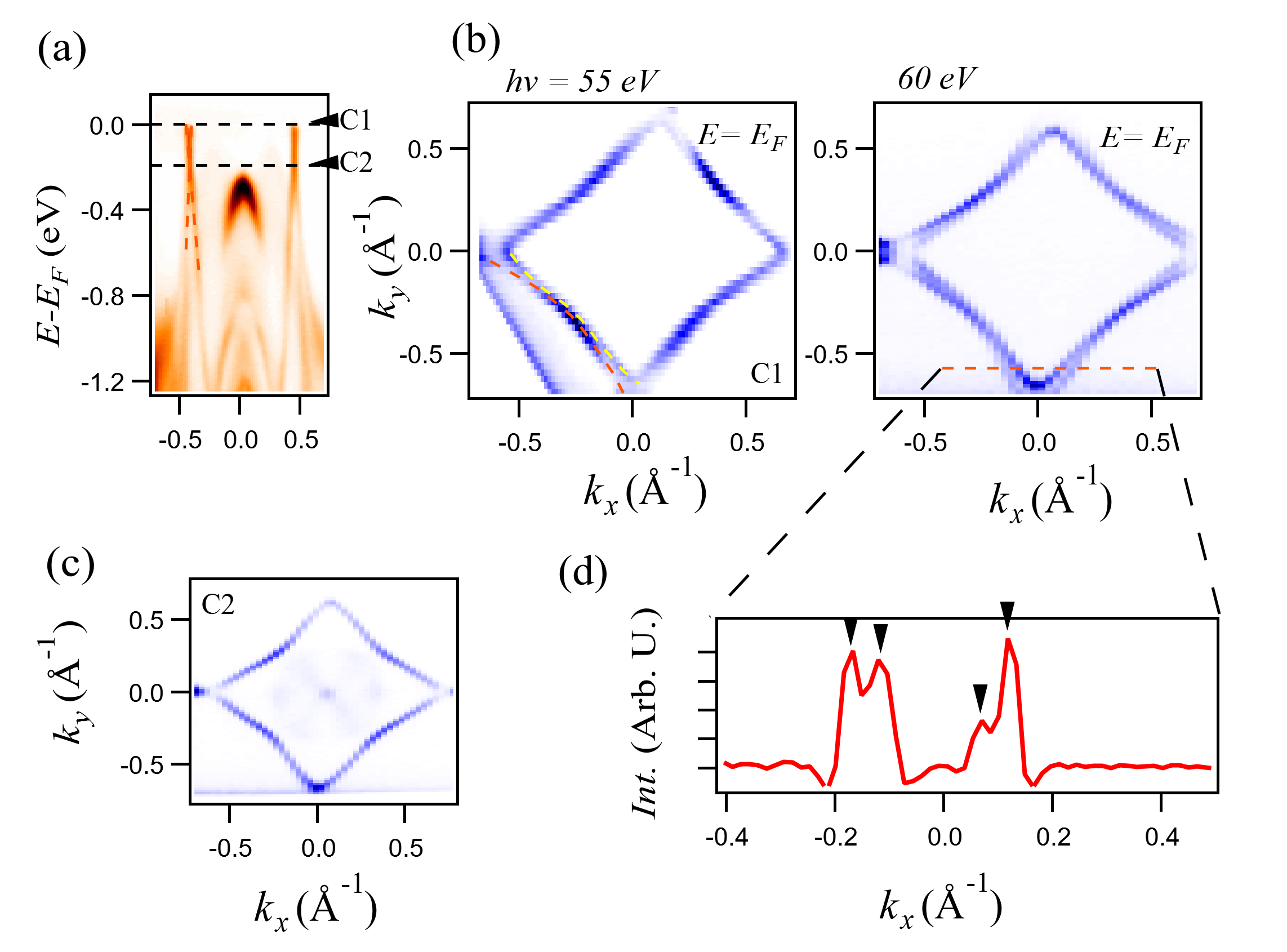}
	\caption{\justifying The double sheet diamond shaped Fermi surface of TbSbTe electronic structure. (a) ARPES cut along the $\overline{\Gamma}-\overline{M}$ high symmetry direction, (b) the Fermi surface cuts with $h\nu$ = 55 and 60 eV (aken at position C1) showing the double sheet nature. (c) CEC at position C2. (d) Intensity profile of the dotted line on (b)(60~eV), exhibiting existence of the double bands.}
  \label{SM13}
\end{figure*}

\begin{figure}[!ht]
	\centering
	\includegraphics[width=15.5 cm]{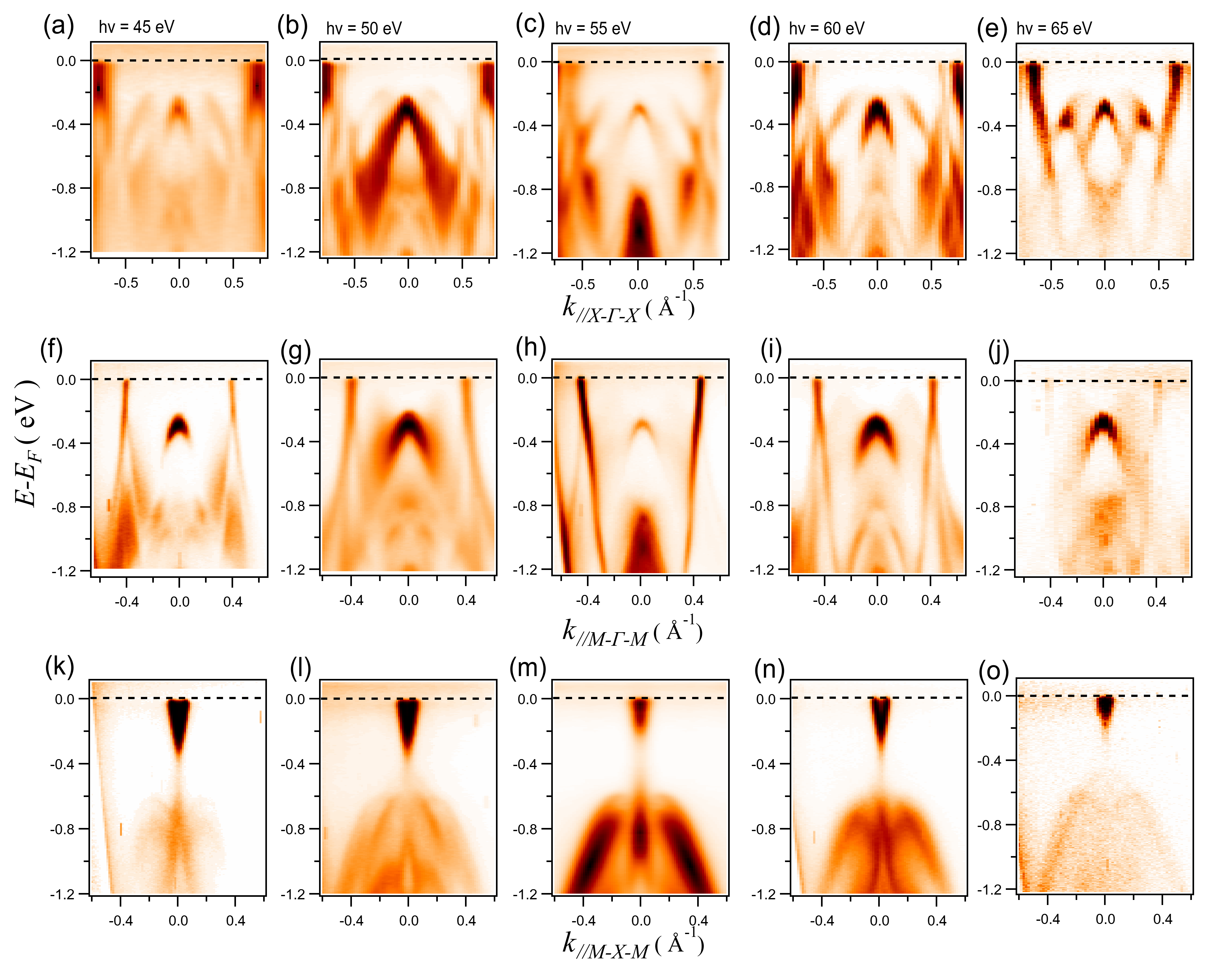}
	\caption
{\justifying Electronic dispersion maps parallel to all the high symmetry directions at different incident photon energies ranging from 45~eV to 65~eV (corresponding to $11.4~\pi/c$ to $13.3~\pi/c$). From (a-e), dispersion maps parallel to $\overline{\text{X}}$-$\overline{\Gamma}$-$\overline{\text{X}}$ are presented. Cuts parallel to $\overline{\text{M}}$-$\overline{\Gamma}$-$\overline{\text{M}}$ and $\overline{\text{M}}$-$\overline{\text{X}}$-$\overline{\text{M}}$ at different incident photon energies are displayed in (f-j) and (k-o) respectively. All the ARPES dispersion maps were collected at beamline 10.0.1.1 in ALS at a temperature of 18~K, corresponding incident photon energies are mentioned in the upper panel of each column.}
  \label{SM_fig_03}
\end{figure} 

\begin{figure}[!hb]
	\centering
	\includegraphics[width=15cm]{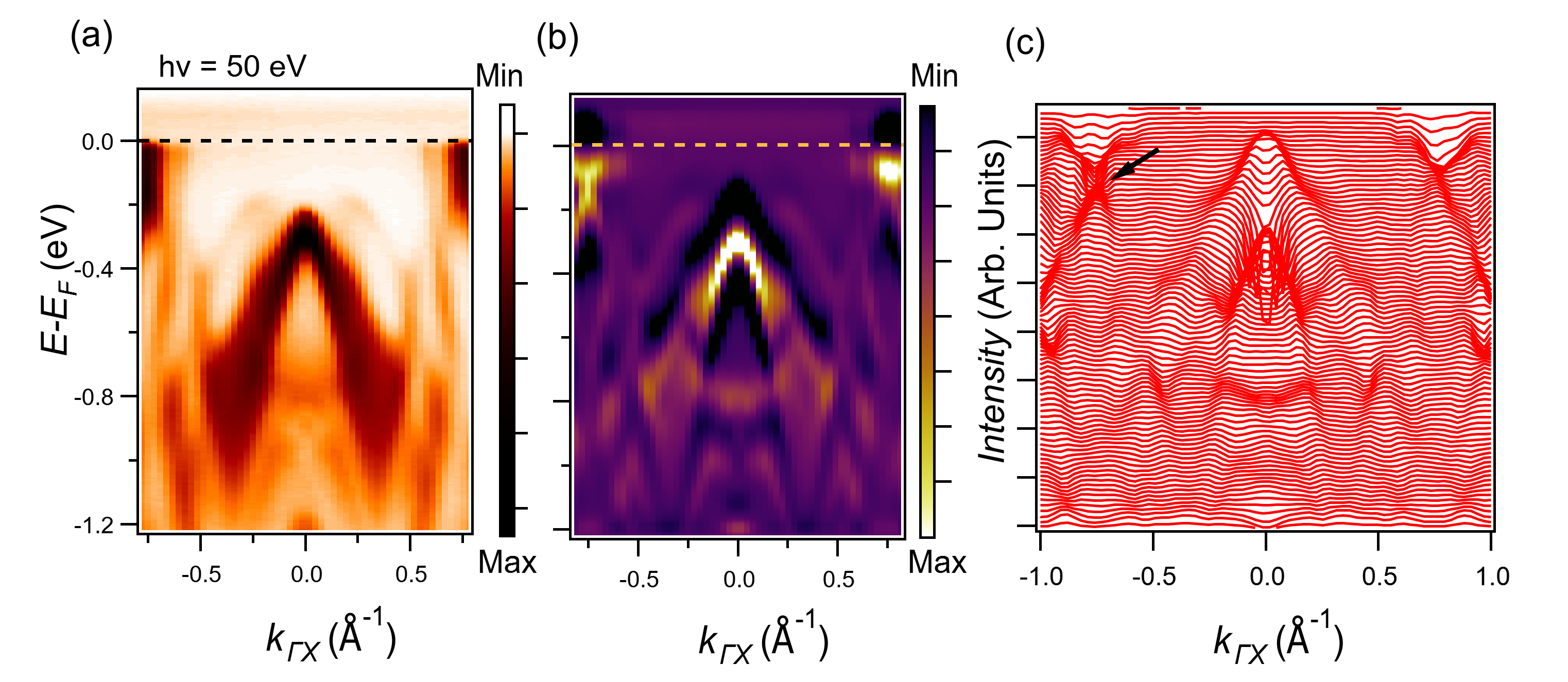}
	\caption{\justifying Observation of Dirac like crossing in bulk bands. (a)
 ARPES map parallel to $\overline{\text{X}}$-$\overline{\Gamma}$-$\overline{\text{X}}$ direction at 50~eV photon energy and (b) its second derivative. (c) Stacked momentum distribution curve (MDC) along $\overline{\text{X}}$-$\overline{\Gamma}$-$\overline{\text{X}}$ exhibiting the band crossing. ARPES data were collected at ALS beamline 10.0.1.1 at a temperature of 18~K.}
  \label{SM5}
\end{figure} 

\begin{figure}[!ht]
	\centering
	\includegraphics[width=14cm]{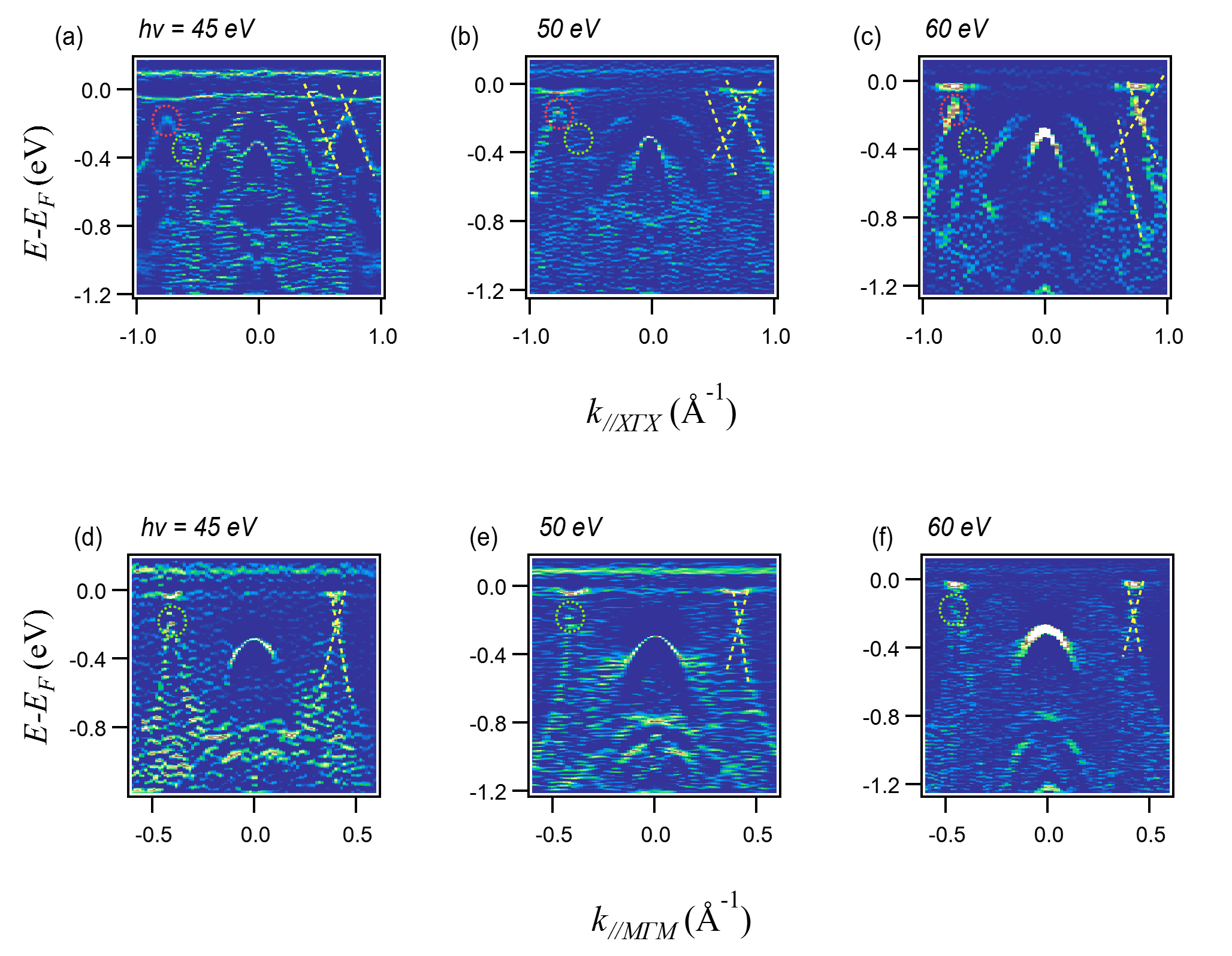}
	\caption{\justifying Position of the Dirac crossings along different HS directions. (a-c) ARPES maps along the $\overline{\text{X}}$-$\overline{\Gamma}$-$\overline{\text{X}}$ and (d-f) high symmetry direction. The broken straight lines act as a guide for eyes, the positions of the crossings are marked with red (nonsymmorphic symmetry protected) and green (not protected by nonsymmorphic symmetry) circles respectively. The ARPES data were collected at ALS beamline 10.0.1.1 at a temperature of 18~K.}
  \label{SM8}
\end{figure} 

\begin{figure}[!hb]
	\centering
	\includegraphics[width=12cm]{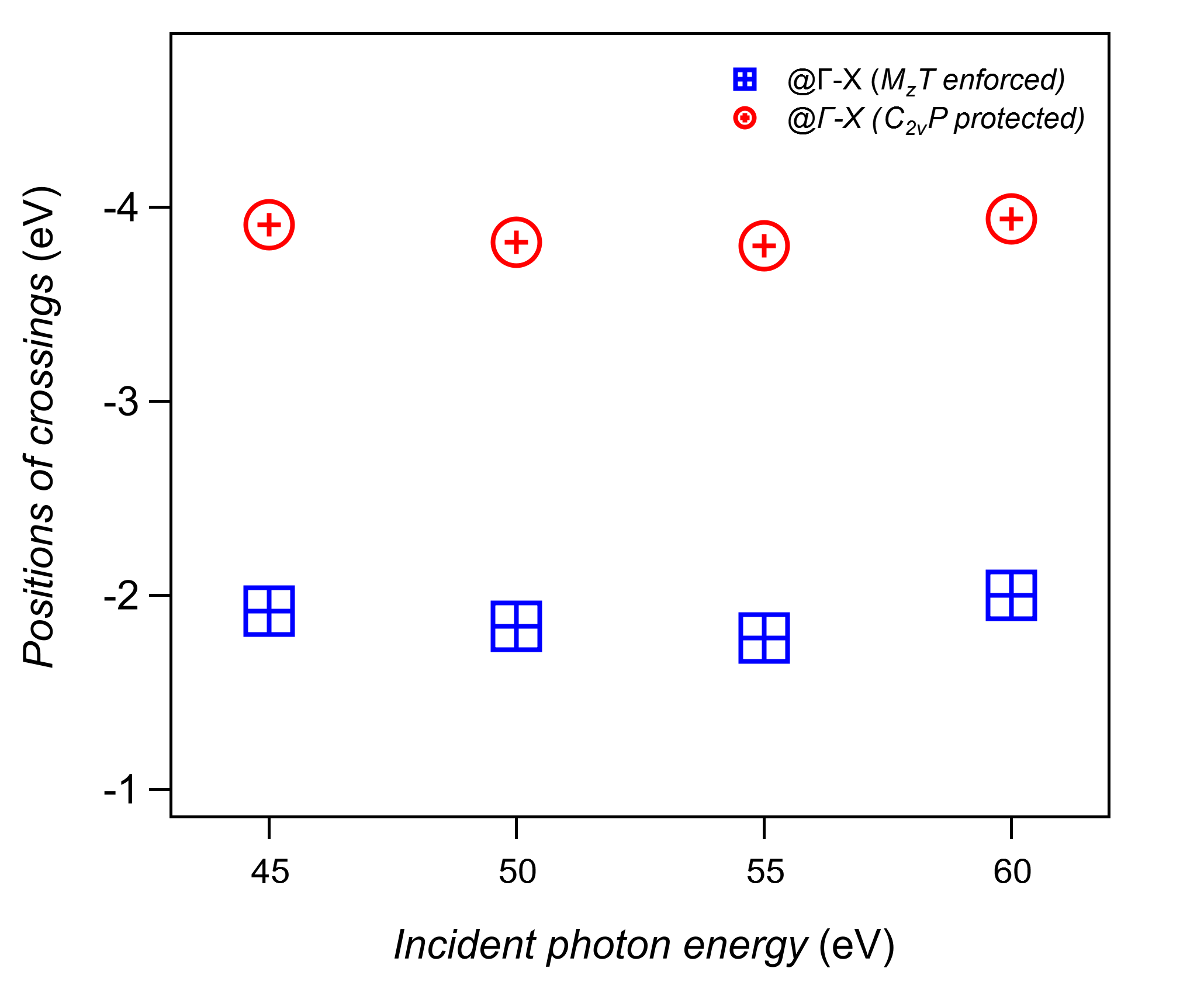}
	\caption{\justifying Positions of the Dirac crossings in bulk bands as a function of incident photon energies.}
  \label{SM9}
\end{figure}

\begin{figure}[!ht]
	\centering
	\includegraphics[width=12cm]{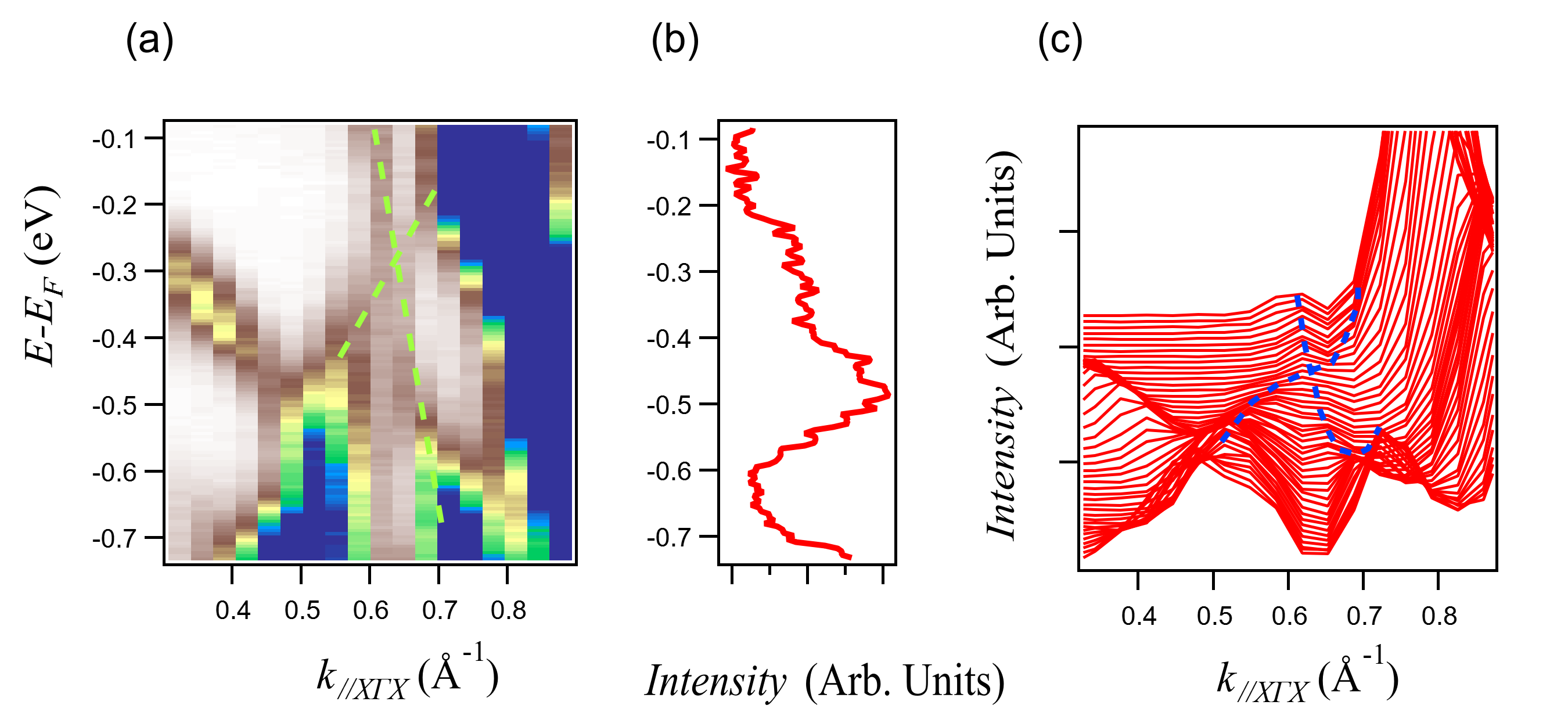}
	\caption{\justifying (a) A magnified view of the Dirac crossing along the $\overline{\Gamma}-\overline{X}$ high symmetry direction (data collected with incident photon energy of 60~eV), (b) EDC taken at the corresponding position of the intersection, (c) stacked MDCs of (a).}
  \label{SM10}
\end{figure}

\end{document}